\documentclass[%
 reprint,
 superscriptaddress,
 amsmath,amssymb,
 aps,
 prx
]{revtex4-2}
\usepackage[T1]{fontenc}
\usepackage{graphicx}% Include figure files
\usepackage{dcolumn}% Align table columns on decimal point
\usepackage{bm}% bold math
\usepackage[colorlinks=true, allcolors=blue]{hyperref}
\usepackage{braket}
\usepackage[table]{xcolor}
\usepackage{xfrac}
\usepackage{comment}
\usepackage{ragged2e}
\usepackage[normalem]{ulem}
\usepackage{mathtools}
\usepackage{layouts}
\usepackage{subcaption}
\usepackage{overpic}
\usepackage{quantikz}
\usepackage{tabularx}
\usepackage{multirow}
\usepackage[section]{placeins}
\usepackage{esint}
\usepackage{enumitem}
\usepackage{amsthm}% for theorems and stuff
\usepackage[capitalise]{cleveref}% load this as late as possible
\newcommand{\up}{\uparrow}
\newcommand{\dw}{\downarrow}

\begin{document}
\title{Single-step high-fidelity three-qubit gates by anisotropic chiral interactions}
\author{Minh T. P. Nguyen}
\email{m.t.phamnguyen@tudelft.nl}
\affiliation{QuTech and Kavli Institute of Nanoscience, Delft University of Technology, Lorentzweg 1, 2628 CJ Delft, The Netherlands}
\author{Maximilian Rimbach-Russ}
\affiliation{QuTech and Kavli Institute of Nanoscience, Delft University of Technology, Lorentzweg 1, 2628 CJ Delft, The Netherlands}
\author{Lieven M. K. Vandersypen}
\affiliation{QuTech and Kavli Institute of Nanoscience, Delft University of Technology, Lorentzweg 1, 2628 CJ Delft, The Netherlands}
\author{Stefano Bosco}
\email{s.bosco@tudelft.nl}
\affiliation{QuTech and Kavli Institute of Nanoscience, Delft University of Technology, Lorentzweg 1, 2628 CJ Delft, The Netherlands}

\newtheorem{theorem}{Theorem}[section]
\newtheorem{corollary}[theorem]{Corollary}
\newtheorem{lemma}[theorem]{Lemma}
\newtheorem{proposition}[theorem]{Proposition}
\newtheorem{criterion}{Criterion}
\newtheorem{appendixcriterion}{Criterion}[section]
\theoremstyle{definition}
\newtheorem{assumption}{Assumption}
\theoremstyle{remark}
\newtheorem*{remark}{Remark}
\theoremstyle{definition}
\newtheorem{definition}[theorem]{Definition}
%\numberwithin{equation}{section} %numbers equations by section, i.e. (sec.eqno)
\Crefname{criterion}{Criterion}{Criteria}
\crefname{criterion}{Crit.}{Crit.}

\begin{abstract}
  Direct multi-qubit gates are becoming critical to facilitate quantum computations in near-term devices by reducing the gate counts and circuit depth. Here, we demonstrate that fast and high-fidelity three-qubit gates can be realized in a single step by leveraging small anisotropic and chiral three-qubit interactions. These ingredients naturally arise in state-of-the-art spin-based quantum hardware through a combination of spin-orbit interactions and orbital magnetic fields. These interactions resolve the key synchronization issues inherent in protocols relying solely on two-qubit couplings, which significantly limit gate fidelity. We confirm with numerical simulations that our single-step three-qubit gate can outperform existing protocols, potentially achieving infidelity $\leq 10^{-4} $ in 80-100 ns under current experimental conditions. To further benchmark its performance, we also propose an alternative composite three-qubit gate sequence based on anisotropic two-qubit interactions with built-in echo sequence and show that the single-step protocol can outperform it, making it highly suitable for near-term quantum processors. 
\end{abstract}

\maketitle

\section{Introduction}
Semiconductor spin qubits are a promising platform for quantum computing due to their long coherence times \cite{Burkard2023,philips2022universal,Zhang2025,hendrickx2021four,Chien-An2024,takeda2022quantum,vandersypen2017interfacing,burkard2023semiconductor,kloeffel2013prospects,stano2022review,scappucci2021germanium,zajac2018resonantly} and are rapidly emerging as a leading platform for scalable quantum computing due to their small footprint and compatibility with industrial semiconductor fabrication techniques \cite{huckemann2024industriallyfabricatedsingleelectronquantum,george202412spinqubitarraysfabricated300,Maurand2016,Jirovec2021,steinacker2024300mmfoundrysilicon,Liles2024,Camenzind2022,Geyer2024,zwerver2022qubits,bosco2023phase,neyens2024probing,xue2021cmos,Weinstein2023}. Experimental advances have demonstrated single-qubit gate fidelities exceeding $99.9\%$ \cite{Yoneda2018,Yang2019} and two-qubit gate infidelities above $99\%$ in mid-scale devices \cite{Xue2022,noiri2022fast,mills2022two,lawrie2023simultaneous,zajac2018resonantly,mkadzik2022precision}. Additionally, progress in coherent links enabling spin coupling over long distances holds promise for large-scale quantum processors with dense connectivity \cite{maxim2024,Zwerver2023,seidler2022conveyor,noiri2022shuttling,langrock2023blueprint,bosco2024high,Dijkema2025,PhysRevX.12.021026,Mi2018,doi:10.1126/science.aar4054,Landig2018,Yu2023,DePalma2024,Janik2025,Burkard2020,PhysRevB.100.081412,PhysRevB.88.241405,PhysRevLett.129.066801}. With high-yield fabrication now achieved over large wafers \cite{koch2024industrial300mmwaferprocessed}, spin qubits are reaching the state where performing quantum computations is becoming increasingly viable \cite{neyens2024probing,huang2024high}.

 Many essential quantum algorithms, including Quantum Machine Learning \cite{Tacchino2019}, block encoding \cite{Guang2017}, Shor’s algorithm, the Quantum Fourier Transform \cite{Nam2020}, and various subroutines, are most naturally expressed using multi-qubit gates, where a unitary operation is applied to a target qubit conditioned on multiple control qubits. While single- and two-qubit gates form a universal gate set, multi-qubit gates are often inefficient to decompose into these primitives \cite{Gheorghiu2022,long2024minimalevolutiontimesfast}.  A notable example is the Toffoli (controlled-controlled-not) gate, a three-qubit gate that requires six two-qubit and nine single-qubit operations when implemented using standard gate sets \cite{shende2008cnotcosttoffoligates}. 
 The single-step resonant Toffoli-like gate \cite{Gullans2019} has been experimentally demonstrated in both silicon \cite{takeda2022quantum} and germanium \cite{hendrickx2021four} spin-qubit platforms. However, the reported gate fidelity remains modest ($\leq 90\%$), primarily because dephasing and additional phase errors caused by off-resonant transitions (ac stark shifts).  For practical near-term applications, the ability to implement fast and high-fidelity multi-qubit gates in a single step would significantly reduce circuit depth, thereby mitigating errors from decoherence and gate imperfections.
\begin{figure}
    \centering
    \includegraphics[width= 0.9 \linewidth]{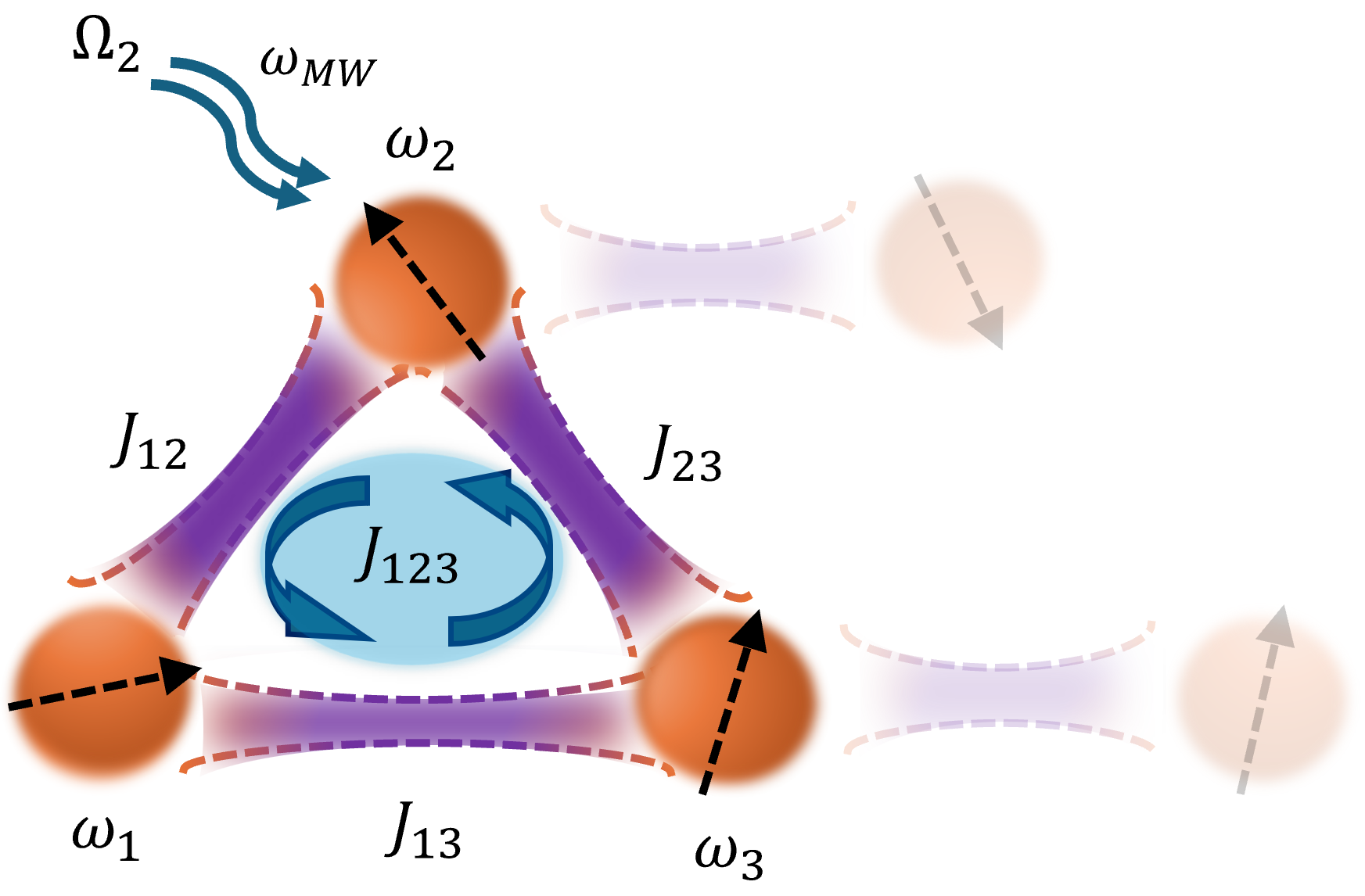}
    \caption{\justifying \textbf{Sketch of a three-qubit setup.} We consider a triangular arrangement comprising three tunnel-coupled QDs each containing a single spin with frequency $\omega_i$. This system displays both nearest-neighbor exchange interactions $J_{ij}$ and  chiral anisotropic interactions $J_{123}$. The target qubit $Q_2$ of our controlled-controlled-rotation gate is driven by a microwave pulse with frequency $\omega_{\rm MW}$ and amplitude $\Omega_2$. 
    }
    \label{fig: triangular-device}
\end{figure}

In this work, we introduce a protocol that enables fast, high-fidelity single-step three-qubit gates in state-of-the-art spin-based hardware. Our approach leverages a small anisotropic chiral multi-qubit interaction to overcome key synchronization issues \cite{Russ2018} between on/off-resonance transitions arising in alternative proposals \cite{Rasmussen2020,Gullans2019,Arias2021}. The protocol enables a consistent improvement of the fidelity by three orders of magnitude compared to using only two-qubit exchange interactions. We demonstrate that this interaction arises naturally in experimentally available state-of-the-art quantum dot (QD) configurations \cite{Chien-An2024,Acuna3holeQ} as sketched in Fig.~\ref{fig: triangular-device}, and emerges from a nontrivial interplay between spin-orbit interactions (SOI) and orbital magnetic fields. Our approach is compatible to both holes and electron spin qubits by intrinsic and engineered SOI using micro and nanomagnets~\cite{scappucci2021germanium,Fang_2023,Burkard2023,unseld2024baseband}. Given that similar multi-qubit interactions have been identified in other qubit platforms \cite{luo2024,Katz2022,Pederson2019,lu2025kramersprotectedhardwareefficienterrorcorrection}, we expect our protocol to be broadly applicable beyond spin qubits. Additionally, we find that the protocol is robust against experimentally relevant sources of error and moderate noise. To further benchmark its performance, we introduce an alternative four-step protocol for the three-qubit gate based on anisotropic two-qubit interactions with a built-in echo sequence \cite{Groenland_2019,Rasmussen2020}. We show that while this approach significantly improves fidelity compared to existing methods, it has in general lower fidelity than the single-step protocol.

The paper is organized as follows. In Section~\ref{section: ideal Hamiltonian}, we briefly review the current approach for implementing single-step three-qubit gates, emphasizing the synchronization issue and its implications on gate fidelity. We then introduce the single-step three-qubit gate protocol that leverages a small three-qubit interaction and resolves this key challenge. Additionally, we propose an alternative four-step protocol based on the echo sequence for architectures that support only two-qubit interactions. In Section~\ref{section: anisotropic chiral interaction}, we discuss the emergence of three-qubit interactions in spin qubit platforms and propose a protocol to characterize them. By using a complete Fermi-Hubbard model including SOI and orbital magnetic fields, we show that these interactions can be engineered in current devices. Finally, in Section~\ref{section: numerical results}, we benchmark the performance of the proposed protocols. By performing numerical simulations using realistic experimental parameters, we demonstrate that even in the presence of large noise, the proposed three-qubit gate achieves a fidelity up to 0.99 at gate times below 100 nanoseconds, outperforming alternative approaches.

\section{Resonant three-qubit gates }
\label{section: ideal Hamiltonian}
\subsection{Target gate}
In this work, we focus on implementing in a single-step a doubly-controlled $\pi$-rotation by the $Y$-axis, i.e. a $\rm C^{2}Ry(\pi)$ gate, and we consider resonant gate protocols~\cite{Gullans2019, Russ2018,zajac2018resonantly}. This gate is a non-Clifford operation, similar to the widely-used Toffoli gate; a fast and high-fidelity implementation of $\rm C^{2}Ry(\pi)$ enables a straightforward realization of the Toffoli gate without additional technical constraints, see e.g. the protocol proposed in Ref.~\cite{Gullans2019}. 

We begin by considering an idealized Ising Hamiltonian consisting solely of Pauli-$Z$ interactions, with no flip-flop terms aside from the applied microwave pulse. This simplification is generally valid when the difference of individual qubit energies is much larger than the exchange interactions and allows us to directly assess the ultimate limit of fidelity for these gates caused by systematic synchronization issues. We will relax this assumption and consider the full interacting system in Sec.~\ref{section: anisotropic chiral interaction} and Sec.~\ref{section: numerical results}. 

We illustrate the device in Fig.~\ref{fig: triangular-device}. This architecture is compatible with current silicon germanium based quantum processors \cite{Chien-An2024,Acuna3holeQ,Sun2024,unseld2024baseband}.
The triple quantum dot system we study is a simple geometry that can exhibit ferromagnetism~\cite{lieb1962ordering,Higgin1950,Elfimov2002}. Its quantum transport properties have been extensively analyzed: the inherent $C_3$ symmetry gives rise to dark states that suppress particle current through the device~\cite{Niklas2017,Kostyrko2009}. In addition, the $S=1/2$ sector of the triple quantum dot has been proposed as a decoherence‐free subspace for encoding a logical qubit, offering built‐in protection against global magnetic field fluctuations~\cite{Scarola2004,Scarola2005,HAWRYLAK2005508}. For a thorough review of the triple quantum dot physics, we refer to Ref.~\cite{hsieh2012physics}.

For concreteness, we now restrict ourselves to the $\rm C^{2}Ry(\pi)$ gate, where qubits $Q_1$ and $Q_3$ serve as control, and $Q_2$ is the target qubit. This means that $Q_2$ undergoes a  $\rm Ry(\pi)$ rotation when $Q_1$ and $Q_3$ are in the state $\ket{1}$. In matrix form, this gate reads
\begin{equation}
    \label{eq: ideal CCRy gate}
    \rm C^2 Ry(\pi) = \begin{bmatrix}
        1 & 0 &  0& 0 & 0 & 0 & 0 & 0  \\
        0 & 1 & 0 &  0 & 0 & 0 & 0 & 0 \\
        0 & 0 & 1 & 0  &0 & 0 & 0 & 0 \\
        0 & 0 & 0 & 1 & 0 & 0 & 0 & 0 \\
        0 & 0 & 0 &0 & 1 & 0 &0 & 0 \\
        0 & 0  & 0 & 0  & 0 & 0 & 0 & -1 \\
        0 & 0 & 0 & 0 & 0 & 0 & 1  & 0 \\
        0 &0 & 0 & 0 & 0 & 1 & 0 & 0
    \end{bmatrix}\ .
\end{equation}
Single-step resonant implementations of this three-qubit gate utilizing only two-qubit interactions are plagued by critical synchronization issues, which substantially limit their performance, as we discuss in the next section.

\subsection{Synchronization issue}
\label{subsection: synchronization issue}
\subsubsection{Ideal system}
Throughout the paper, we use the unit scale $\hbar =1$. We start from the ideal Ising Hamiltonian describing the triangular setup in Fig.~\ref{fig: triangular-device} and containing only two-qubit interactions
\begin{equation}
    \label{eq: ideal H two-body}
    H_{\rm 2Q} = \sum_{i=1}^{3} \frac{\omega_i}{2} Z_i + \frac{J_{12}^{\parallel}}{4}Z_1 Z_2 + \frac{J_{23}^{\parallel}}{4} Z_2 Z_3 + \frac{J_{13}^{\parallel}}{4} Z_1 Z_3 \ .
\end{equation}
Here, $\omega_{i}$ is the frequency of qubit $Q_i$, assumed to be different for each qubit, and $J^{\parallel}_{ij}$ is the component of the two-qubit exchange interaction between  $Q_i$ and $Q_j$ parallel to the qubit quantization axis \cite{Geyer2024}. We note that this Hamiltonian also describes linear chains of qubits when one of the interactions $J_{ij}^{\parallel}=0$. Individual interactions $J_{ij}$ can be electrically controlled with high precision in current spin-qubit experiments \cite{Weinstein2023,Chien-An2024,Acuna3holeQ,Sun2024}. 

To implement the resonant gate, we consider that the interactions $J^{\parallel}_{ij} $ are turned on for a time $T_{g}$ and during this gate time a single microwave pulse is applied to $Q_2$, resulting in the driving Hamiltonian 
\begin{equation}
    H_{\rm drive}(t) = \Omega_{2} \sin(\omega_{\rm MW} t)~  X_2,
\end{equation}
where $\Omega_{2}$ is the Rabi-frequency and $\omega_{\rm MW}$ the microwave pulse 
frequency. We assume that the Zeeman energy differences between the qubits are sufficiently large, ensuring that any unintended crosstalk where $Q_1$ and $Q_3$ are also inadvertently driven by the microwave pulse, remains far off-resonant. The qubit frequencies differences and the exchange interaction $J^{\parallel}_{ij}$ lift the degeneracy of the transition energies, such that we can target the transition $\ket{101} \leftrightarrow\ket{111}$ by setting the microwave frequency $\omega_{\rm MW}$ to be
\begin{equation}
    \label{eq: MW frequency}
    \omega_{\rm MW } = \omega_{2} - \frac{J_{12}^{\parallel} + J_{23}^{\parallel}}{2}. 
\end{equation}

We remove the time dependence of the qubit frame Hamiltonian  $H_{\rm QF}(t) =H_{\rm 2Q} + H_{\rm drive}(t)$ by moving to the rotating frame defined by the unitary transformation 
\begin{equation}
    U_{\rm RW}(t) = \exp\Big(i  \frac{\omega_{\rm MW} t }{2} \sum_{i} Z_i \Big)
\end{equation}
and performing the rotating wave approximation to obtain 
\begin{subequations}
    \label{eq: rotating wave hamiltonian}
    \begin{align}
            & H_{\rm RW}  = U_{\rm RW }(t)H_{\rm QF}(t)  U_{\rm RW}^{\dagger}(t) + i \frac{d U_{\rm RW}(t)}{dt} U_{\rm RW}^{\dagger}(t) \\
             & \approx  \sum_{i} \frac{\omega_i -\omega_{\rm MW}}{2} Z_i + \frac{\Omega_2}{2} Y_2 + \sum_{\langle ij \rangle} \frac{J_{ij}^{\parallel}}{4} Z_i Z_j \ . 
    \end{align}
\end{subequations}
From the approximate Hamiltonian $H_{\rm RW}$, it follows that the $\rm C^{2}Ry(\pi)$ gate is implemented  when the gate time $T_{g}$ satisfies 
\begin{equation}
    \label{eq: timing condition}
    \Omega_2 T_{g} = (4m+1)\pi.
\end{equation}
Here, we introduce the non-negative integer $m$ to also allow for additional trivial $2\pi$ rotations for $Q_2$. The resulting gate obtained in the rotating frame is unitarily equivalent to the ideal $\rm C^2Ry(\pi)$ gate defined in Eq.~\eqref{eq: ideal CCRy gate}, up to the single-qubit phase corrections 
 \begin{equation}
     \hat{U}_{\rm cor} = \exp\Big[ i T_g \sum_{i=1,3} (\omega_i-\omega_{MW}) \frac{Z_i}{2} \Big],
 \end{equation}
and a two-qubit phase correction $\exp(i T_{g} {J_{13}^{\parallel}} Z_1 Z_3/4) $ that compensates for the coherent rotation between $Q_1$ and $Q_3$ induced by $J_{13}^{\parallel}$.

\subsubsection{Synchronization}
Even in this ideal case, however, this gate implementation does not enable perfect fidelity. There are systematic errors originating from a close-in-frequency transitions. While driving the resonant transition $\ket{101} \leftrightarrow \ket{111}$, we also inadvertently drive the other off-resonant transitions $\ket{001} \leftrightarrow\ket{011}$, $\ket{100} \leftrightarrow \ket{110}$, and $\ket{000} \leftrightarrow \ket{010} $. These off-resonant transitions are the key limiting factors for achieving fast and high-fidelity three-qubit gates, as they require a low Rabi frequency $\Omega_2 \ll J^{\parallel}_{12},J^{\parallel}_{23}$ or complex pulse shaping techniques \cite{ustun2024single,spiteri2018quantum,Zahedinejad2015,waldherr2014quantum} to suppress them.

For two-qubit gates, Refs.~\cite{Russ2018,heinz2021crosstalk} show that this issue can be mitigated by tuning the ratio between Rabi frequency and the two-qubit exchange interaction to ensure that all off-resonant transitions undergo a trivial $2\pi$ rotation at the end of the gate. This approach, often called synchronization, is widely used in experiments aiming to reach high-fidelity fast gates. Synchronization, however, cannot be generalized to three-qubit controlled-controlled-rotation gate \cite{heinz2021crosstalk,rimbach2023simple}, and constitutes the main source of infidelity for these gates. 

More explicitly, the off-resonant transitions have no effect on the $\rm C^{2} Ry(\pi)$ gate when
 \begin{subequations}
    \label{eq: sync 2-body}
     \begin{align}
         \label{eq: sync 2-body a}
         \sqrt{(J_{23}^{\parallel})^2 + \Omega_2^2 }~T_{g} & = 4n_1 \pi, \\
          \label{eq: sync 2-body b}
         \sqrt{(J_{12}^{\parallel})^2 + \Omega_2^2 }~T_{g} & = 4n_2 \pi, \\ 
          \label{eq: sync 2-body c}
         \sqrt{(J_{12}^{\parallel} + J_{23}^{\parallel})^2 + \Omega_2^2 }~T_{g} & = 4 n_3 \pi ;
     \end{align}
 \end{subequations}

The tuples of integers $\{m,n_1,n_2,n_3\}$ satisfying Eqs.~\eqref{eq: timing condition} and \eqref{eq: sync 2-body} define the synchronization condition. We note that there is a  physical constraint $m < \min(n_1,n_2,n_3)$ arising from the fact that off-resonant Rabi frequencies are larger than resonant Rabi frequency. If we use the synchronization condition $J_{12}^{\parallel} = J_{23}^{\parallel} = \sqrt{15}~ \Omega_2$, as done in Refs.~\cite{Yoneda2018,Russ2018,Geyer2024,Gullans2019}, we find that Eqs.~\eqref{eq: sync 2-body a} and \eqref{eq: sync 2-body b} are satisfied when $n_1=n_2=1$, but Eq.~\eqref{eq: sync 2-body c} has no solution. This no-go result is more general: no integer tuple $\{m,n_1,n_2,n_3\}$ can satisfy all the synchronization conditions for the $\rm C^{2}Ry(\pi)$ gate.

This issue, which we refer to as the \textit{synchronization issue}, was recognized in Ref.~\cite{Gullans2019} and is general to controlled-controlled-gates, including $\rm C^2 Z$ gates \cite{Jiaan2024}. Intuitively, the synchronization issue can be understood by the simple consideration that in the system of equations of Eq.~\eqref{eq: sync 2-body}, there are four transitions to be synchronized, but only three free parameters $\{ \Omega_2, J_{12}^{\parallel}, J_{23}^{\parallel} \}$. We emphasize that we only have three free parameters instead of four since the interaction $J_{13}^{\parallel}$ is irrelevant to the gate dynamic induced by $ H_{\rm drive}$, and its deterministic conditional-phase operation can be corrected after the gate, as discussed previously. We provide a detailed proof of the synchronization issue for general controlled-controlled-rotation gate implemented with only two-qubit interaction in Appendix.~\ref{appendix: proof of infeasiblity}.

\subsubsection{Impact on fidelity}

The synchronization issue cannot be fully resolved by using only two-qubit interactions and it substantially limits the fidelity of fast three-qubit gates. We now show that in unsynchronized gates present an unavoidable trade-off between gate speed and fidelity. 

For simplicity, we assume identical couplings $J_{12}^{\parallel} = J_{23}^{\parallel} = J^{\parallel}$ between $Q_1 \leftrightarrow Q_2$ and $Q_2 \leftrightarrow Q_3$. Since it is impossible to satisfy all synchronization conditions simultaneously, we prioritize fulfilling the first three [Eqs.~\eqref{eq: timing condition},~\eqref{eq: sync 2-body a}, and \eqref{eq: sync 2-body b}], which correspond to the on-resonant and first-order off-resonant transitions. These three synchronization equations are parameterized by a pair of integers $(m,n)$ 
\begin{subequations}
    \begin{align}
        \Omega_2~T_{g} & = (4m+1)\pi, \\ 
        \sqrt{ (J^{\parallel})^2 + \Omega_2^2 }~T_{g} & = 4 n \pi. 
    \end{align}
\end{subequations}
As a result, the gate error arises from the second-order off-resonant transition $\ket{000}\leftrightarrow\ket{010}$. While other synchronization pairs from Eq.~\eqref{eq: sync 2-body} can be chosen, for instance Eqs.~\eqref{eq: sync 2-body a} and \eqref{eq: sync 2-body c}, such choices would lead to lower gate fidelity, as the resulting unsynchronized transition would occur at the first-order off-resonant level rather than the second-order.

For a chosen pair $(m,n)$, we define the average fidelity of the $\rm C^2 Ry(\pi)$ gate in the rotating frame, including qubit phase correction $\hat{U}_{\rm cor}$,  to be 
\begin{equation}
\bar{F}(m,n)  =  \int d\mu (\psi) |\bra{\psi}  \rm{C^2Ry}^{\dagger}(\pi)  \hat{U}_{\rm{cor}}~e^{-i T_{g} H_{\rm RW}}     \ket{\psi}|^2, 
\end{equation}
where $d \mu(\psi)$ is the random state Haar-measure. The average gate fidelity $\bar{F}(m,n)$ can be computed analytically using the entanglement fidelity \cite{NIELSEN2002249}. For a fixed exchange interaction $J^{\parallel}$, the average gate fidelity $\bar{F}(m,n)$ and gate duration $T_{g}$ are given by 
\begin{subequations}
    \label{eq: two-body gate fidelity and duration}
    \begin{align}
        \bar{F}(m,n) & =  \frac{8}{9}\left|\frac{3}{4} + \frac{1}{4}\cos\Big(4\pi n \sqrt{ 1 - \frac{3(4m+1)^2}{64 n^2}   }  \Big)  \right|^2+\frac{1}{9}, \\
        T_{ g} & = \frac{\pi}{J^{\parallel}} \sqrt{16n^2 - (4m+1)^2}.
    \end{align}
\end{subequations}  
We observe that the average fidelity $\bar{F}(m,n)$ is independent of the absolute values of the exchange interaction $J^{\parallel}$ and the Rabi frequency $\Omega_2$, assuming no decoherence; instead, it depends only on their ratio $J^{\parallel}/\Omega_2$ indirectly through the chosen synchronization pair $(m,n)$. Consequently, reducing $\Omega_2$ alone does not improve the gate fidelity if $(m,n)$ remains fixed. 

Closer inspection of Eqs.~\eqref{eq: two-body gate fidelity and duration} reveals a trade-off: the fastest  $\rm C^{2}Ry(\pi)$ gates have the lowest fidelity. For instance, if $n$ is fixed, increasing $m$ shortens the gate time $T_{g}$ but also significantly reduces fidelity. Specifically, maximizing fidelity for a given large $n$ requires setting $m=0$,  in which case the average gate fidelity $\bar F(0,n)$ and the gate duration $T_{g}$ are given by 
\begin{subequations}
    \begin{align}
        \label{eq: infidelity synchronization}
        1-\bar{F}(0,n) & \approx \frac{\pi^2}{512} \frac{1}{n^2}, \\
        \label{eq: gate time with n}
        T_{g} & \approx \frac{4 \pi}{J^{\parallel}} n.
    \end{align}
\end{subequations}
This result implies that the gate infidelity decreases by a factor of four each time the gate duration $T_{g}$ is doubled. For example, choosing $(m,n)=(0,1)$,  which corresponds to the fastest $\rm C^{2}Ry$ gate for a fixed exchange interaction $J^{\parallel}$, yields a gate fidelity of approximately $\bar{F}(0,1) \approx 0.98$. To reach four-nines fidelity ($\bar{F}(0,16) \approx 0.9999$) we can select $(m,n)=(0,16)$, but at the cost of increasing the gate duration by a factor of 16.

For these slow gates, however, the ideal Ising Hamiltonian implicitly assumed in Eq.~\eqref{eq: infidelity synchronization} also becomes inaccurate because of the neglected flip-flop terms, as we will show in Sec.~\ref{section: numerical results}. As a result, in real experiments, increasing the gate time $T_{g}$ does not improve fidelity, but rather drastically decreases it because the additional exchange dynamics introduces systematic errors that quickly become the limiting factor for the fidelity of direct three-qubit gates. We now show how to circumvent this trade-off and achieve fast and high-fidelity gates by introducing a small transverse $ZZZ$ interaction.

\subsection{Ideal gate with three-qubit interaction}
As discussed above, we cannot implement the $\rm C^2Ry(\pi)$ gate perfectly in a single-step using only two-qubit interactions. However, if  spin qubits are arranged in a triangular loop, as for example sketched in Fig.~\ref{fig: triangular-device}, the system can present a small three-qubit interaction. We delay the discussion of the physical requirement for these interactions to appear to Sec.~\ref{section: anisotropic chiral interaction} and focus here on the effect of diagonal three-qubit interactions on resonant three-qubit gates. We consider the Hamiltonian
\begin{equation}
    \label{eq: ideal Hamiltonian three body}
    H_{\rm loop }  = H_{\rm 2Q}  + \frac{J_{123}^{\parallel}}{8} Z_1 Z_2 Z_3 , 
\end{equation}
which comprises the ideal Ising Hamiltonian in Eq.~\eqref{eq: ideal H two-body} and an additional three-qubit $ZZZ$ interaction.

Surprisingly, we find that even a small three-qubit interaction $J_{123}^{\parallel} \ll J_{12}^{\parallel},J_{23}^{\parallel}$ is sufficient to resolve the synchronization issue, enabling exact three-qubit gates in a single step.
While here we focus on QD spin qubits, we emphasize that tunable many-body interactions emerge in various qubit platforms, including cold atom systems \cite{luo2024}, trapped ions \cite{Katz2023}, superconducting circuits \cite{Pederson2019}, and Andreev spin qubits \cite{lu2025kramersprotectedhardwareefficienterrorcorrection}. \\
The single-step gate protocol for the $\rm C^{2}Ry$  gate remains the same as discussed in Sec.~\ref{subsection: synchronization issue} but with the modified microwave field frequency 
\begin{equation}
    \omega_{\rm MW} = \omega_{2} - \frac{J_{12}^{\parallel} + J_{23}^{\parallel}}{2} + \frac{J_{123}^{\parallel}}{4}.
\end{equation}
The new synchronization conditions for $H_{\rm loop }$ are 
 \begin{subequations}
    \label{eq: synchronization condition 3Q}
     \begin{align}
        \Omega_2 ~ T_{ g} & = (4m+1)\pi\\
        \sqrt{\Big(J_{23}^{\parallel} - \frac{J_{123}^{\parallel}}{2} \Big)^2 + \Omega_2^2 }~T_{ g} & = 4n_1 \pi, \\
        \sqrt{\Big(J_{12}^{\parallel} - \frac{J_{123}^{\parallel}}{2} \Big)^2 + \Omega_2^2 }~T_{ g} & = 4n_2 \pi, \\ 
        \sqrt{(J_{12}^{\parallel} + J_{23}^{\parallel})^2 + \Omega_2^2 }~T_{ g}& = 4 n_3 \pi. 
     \end{align}
 \end{subequations}
The additional degree of freedom provided by $J_{123}^{\parallel}$ enables exact synchronization of the gate for any chosen tuple $(m,n_1,n_2,n_3)$ with $m < n_i$. 
%to tune the Rabi frequency $\Omega_2$ and exchange interactions $\{ J^{\parallel}_{ij}, J^{\parallel}_{123} \}$ to satisfy the synchronization conditions for any chosen tuple $(m,n_1,n_2,n_3)$ with $m < n_i$.

For example, assuming identical exchange couplings $J_{12}^{\parallel} = J_{23}^{\parallel} = J^{\parallel}$ and selecting $(m,n_1,n_2,n_3) = (0,1,1,2)$, the coupling coefficient ratios $J^{\parallel},J^{\parallel}_{123}$ and the Rabi frequency $\Omega_2$ are given by 
\begin{subequations}
    \label{eq: synchronization ratios}
    \begin{equation}
        \frac{J^{\parallel}}{\Omega_2} = \frac{3\sqrt{7}}{2} \approx 3.97
    \end{equation}
    \begin{equation}
        \frac{J_{123}^{\parallel}}{\Omega_2} = 3\sqrt{7} - 2\sqrt{15}   \approx 0.19. 
    \end{equation}
\end{subequations}
We emphasize that these ratios are physically realistic, and within experimental reach, as three-qubit interactions are typically an order of magnitude smaller than two-qubit interactions \cite{Sen1995}. We select the tuple $(0,1,1,2)$ for the proposed three-qubit gate, as it yields the fastest gate implementation with realistic parameters. More generally, for tuples of the form $(0,n,n,2n)$, the ratio $J^{\parallel}_{123}/J^{\parallel}$  can be further reduced to an arbitrary level by increasing $n$, thereby relaxing hardware constraints. For completeness, we provide the synchronization ratios for arbitrary tuples $(m,n,n,2n)$ in Appendix.~\ref{appendix: proof of infeasiblity}. 

Notably, when the synchronization conditions in Eqs.~\eqref{eq: synchronization condition 3Q} are satisfied, the fidelity of the $\rm C^2Ry(\pi)$ gate is equal to 1, regardless of the chosen tuple $(m,n_1,n_2,n_3)$, in stark contrast to the $J_{123}^\parallel=0$ case.

In this section, we demonstrated that introducing a small three-qubit $ZZZ$  interaction can resolve the synchronization issue in the implementation of the $\rm C^2Ry(\pi)$ gate. However, we emphasize that this approach is not limited to this specific gate. As shown in Ref.~\cite{Jiaan2024}, an ideal $\rm C^2 Z$ gate can also be realized using an additional $ZZZ$ interactions. More broadly, such three-body interactions enable the implementation of a wide class of three-qubit gates that act non-trivially on a designated two-dimensional subspace of the full eight-dimensional Hilbert space, while leaving the orthogonal subspace invariant.

\subsection{Four-step echo-protocol}
\begin{figure}
    \centering
    \includegraphics[width=\linewidth]{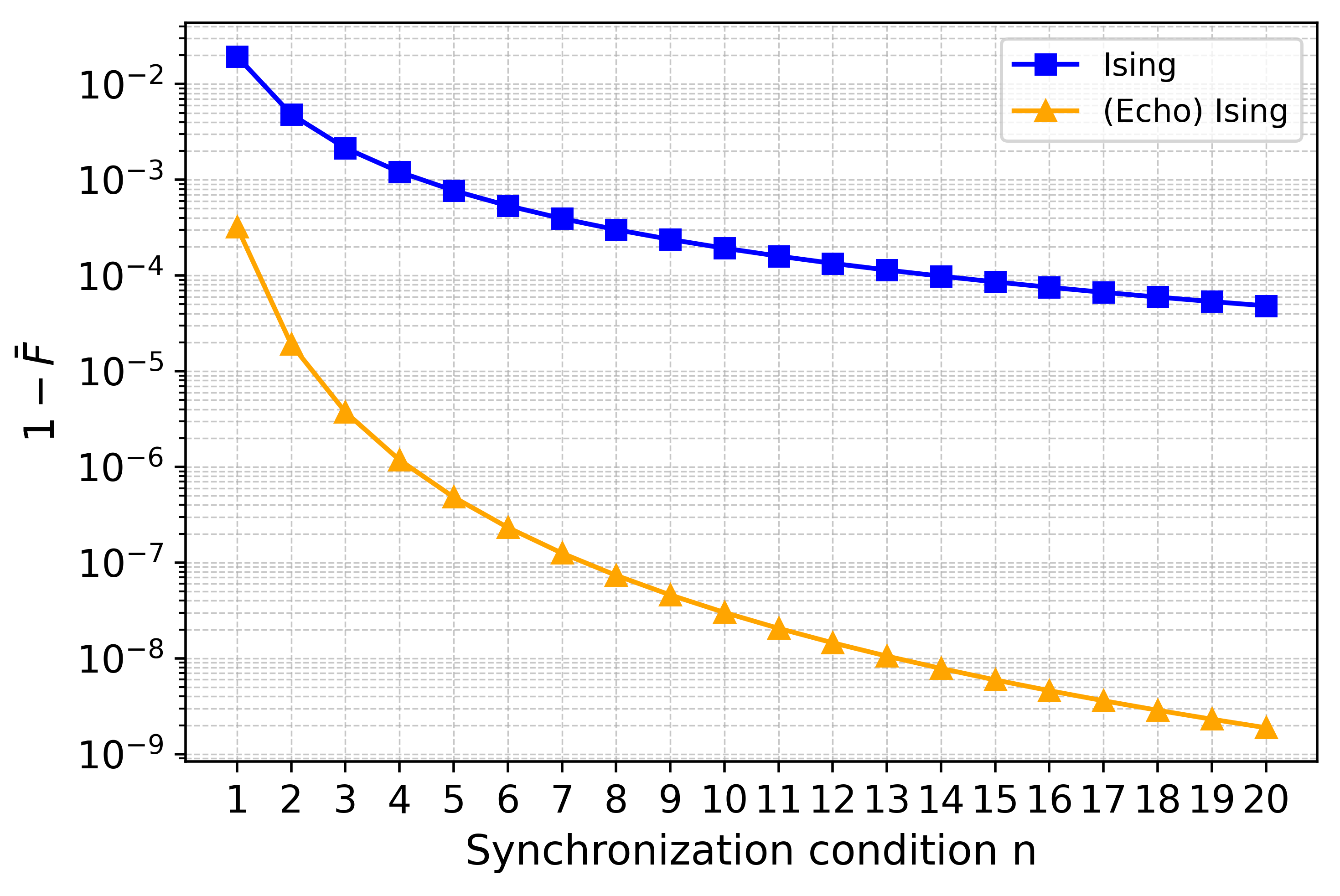}
    \caption{\justifying \textbf{Average infidelity of the $\rm C^2 Ry(\pi)$ gate for increasing gate time when relying on two-qubit interactions only.} The gate time is parameterized by the synchronization condition $n$, where $n={T_g J^{\parallel}}/{4\pi}$. We consider two-qubit gate using Ising interaction (blue) and its Echo variant (yellow). We restrict ourselves to systematic synchronization errors, which dominate for fast gates with small $n$. The anisotropic gate follows a $1/n^2$ scaling, as described by Eq.~\eqref{eq: infidelity synchronization}, while its Echo variant follows a $1/n^4$ scaling, as given by Eq.~\eqref{eq: infidelity echo}. We also emphasize that with the assumptions used here the $\rm C^2 Ry(\pi)$ gate has fidelity of 1, independent of $n$, when utilizing the $ZZZ$ interaction.  }
    \label{fig: ideal gate infidelity}
\end{figure}
As an alternative approach to improve the fidelity of fast three-qubit gates in physical systems supporting only two-qubit interactions, we propose here a four-step echo protocol. This approach requires two additional single-qubit gates compared to the fully synchronized three-qubit interaction-based protocol, but can enhance the gate fidelity by orders of magnitude compared to the conventional $J_{123}^\parallel=0$ case. 

As discussed below Eq.~\eqref{eq: two-body gate fidelity and duration}, we identify the main source of infidelity to be the second-order off-resonant transition $\ket{000} \leftrightarrow \ket{010}$.  Within this $2\times2$ subspace, the time evolution corresponds to a free precession with a large $Z$-axis frequency $J_{12}^{\parallel} + J_{23}^{\parallel}$ and a small $Y$-axis frequency $\Omega_2$. This unwanted precession can be effectively counteracted if the sign of the $Z$-axis frequency in this subspace is flipped precisely at the midpoint of the time evolution \cite{Groenland_2019,Rasmussen2020}. 

In practice, implementing such a sign flip requires additional control overhead and may even be infeasible on certain platforms, such as spin qubits, where the exchange interaction is always positive. We show, however, that it is possible to achieve the sign flip at the circuit level by mimicking an echo sequence \cite{Russ2018}.  

Here, we use the Pauli phase-tracking frame implemented in software, meaning that all single-qubit phase gates are virtual gates. Within this frame, the echo variant of the single-step $\rm C^2 Ry(\pi)$ gate is given by 
\begin{equation}
    \label{eq: Hahn Echo structure}
    Y_2 \exp\Big(-i \frac{T_{g}}{2} H_{\rm RW}  \Big) Y_2 \exp\Big(-i \frac{T_{g}}{2} H_{\rm RW}  \Big),
\end{equation}
where $H_{\rm RW}$ is the Hamiltonian in Eq.~\eqref{eq: rotating wave hamiltonian}. The Pauli-$Y$ gate required for the echo process is applied to the target qubit $Q_2$. This echo sequence can be readily adapted to other controlled-controlled-rotation gates by appropriately modifying the echo pulse.

To analytically demonstrate that the protocol suppresses the residual synchronization error, we decompose the time evolution into the block-diagonal form 
\begin{subequations}
    \begin{equation}
    e^{-i \frac{T_{g}}{2} H_{\rm RW}} = \oplus_{s_1 s_3} e^{ -i \frac{T_{g}}{2} \frac{\sqrt{L_{s_1 s_3}^2 + \Omega_2^2}}{2} ~ \vec{n}_{s_1 s_3} \cdot \vec{\sigma}   },
    \end{equation}
    \begin{equation}
        L_{s_1 s_3} = \bar{s}_1 J_{12}^{\parallel} + \bar{s}_3 J_{23}^{\parallel},
    \end{equation}
    \begin{equation}
        \vec{n}_{s_1,s_3} = \frac{(0,\Omega_2, L_{s_1 s_3}  )}{\sqrt{(L_{s_1 s_3})^2 + \Omega_2^2 }} .
    \end{equation}
\end{subequations}
The binary indices $s_{1},s_{3} \in \{0,1 \}$  indicate the states of the control qubits $Q_1$ and $Q_3$, $\bar{s}_i$ is the negation of $s_i$, $L_{s_1,s_3}$ represents the energy gap between the states $\{\ket{s_1, 1, s_3 }$ and $\ket{s_1, 0, s_3 }\}$, and $\vec{n}_{s_1,s_3}$ is the free precession axis of that subspace. 

The effect of Pauli-$Y$ conjugation on the second time-evolution block of Eq.~\eqref{eq: Hahn Echo structure} is to flip the $Z$-component $L_{s_1,s_3} \to - L_{s_1 s_3}$ of the  precession axis $\vec{n}_{s_1,s_3}$ in each subspace while keeping the $Y$-component $\Omega_2$ fixed. In the resonant subspace $s_1 = s_3 = 1$, where $L_{1,1,}= 0$, the echo sequence has no effect and we have performed the desired $\rm Ry(\pi)$ rotation. However, in the off-resonant subspaces, the echo structure in Eq.~\eqref{eq: Hahn Echo structure} effectively cancels the free-precession error. Assuming again that  $J_{12}^{\parallel} = J_{23}^{\parallel} = J^{\parallel}$ and choosing the synchronization condition $(m,n)=(0,n)$, it follows directly that the average fidelity is
\begin{subequations}
    \begin{equation}
    \bar{F}_{\rm Echo}(0,n)  = \frac{ 8}{9} \left|1 - \frac{1- \cos(\frac{\pi}{2}\sqrt{64n^2 - 3})  }{4(64n^2-3)}  \right|^2 + \frac{1}{9} ,
    \end{equation}
    \begin{equation}
        \label{eq: infidelity echo}
        1 - \bar{F}_{\rm Echo}(0,n) \approx \frac{\pi^2}{72  } \frac{3}{64n^4 } .
    \end{equation}
\end{subequations}

This result demonstrates that by combining frequency synchronization with an echo-pulse sequence \cite{Russ2018}, we can suppress errors to fourth order in $n$ instead of only quadratically as in Eq.~\eqref{eq: infidelity synchronization}. Furthermore, using the echo gate protocol, we can achieve a fidelity of $0.9999$ at $n=2$ (i.e with double gate time) whereas the single-step protocol reaches only $\bar{F}(0,2) \approx 0.995$ in approximately the same gate time, see Fig.~\ref{fig: ideal gate infidelity}. In practice, the four-step echo protocol requires only one additional single-qubit gate since the last echo gate can be tracked in software, or incorporated in the next pulses applied on the qubit. As a result, we note that this protocol can be a practical approach for implementing multi-qubit gates in a wide range of near-term quantum processors even without perfect synchronization. 

For completeness, in Appendix.~\ref{appendix: generalization to N-qubit}, we  extend the echo protocol to more general $\rm C^{N-1}Ry(\pi)$  gates between $N$-qubits, with $N > 3$. Surprisingly, we find that the general $\rm C^{N-1}Ry(\pi)$ gate has comparable gate-fidelity and gate-time with respect to the three-qubit gate, and even improves as the number of control qubits increases. 

\subsection{Systematic errors from flip-flop and flop-flop terms}
\label{section: systematic errors}
Up to this point, we have assumed that the system Hamiltonian is diagonal by neglecting $X_i X_j$ and $Y_i Y_j$ terms in the two-qubit exchange interactions, relying on the assumption that the qubit frequency differences $|\omega_{i}-\omega_{j}|$ are larger than $J_{ij}$.  In this subsection, we benchmark the proposed gate protocols in the presence of systematic errors arising from these flip-flop terms. 

We classify the single-step gate protocol as follows. We refer to the gate as \textbf{Iso}  and \textbf{Ani} gate if it is implemented using two-qubit exchange interactions, where we have flip-flop terms ${J_{ij}}(X_i X_j + Y_i Y_j)/4$  for the Iso gate and flop-flop terms ${J_{ij}}(X_i X_j -Y_i Y_j)/4$ for the Ani gate. The \textbf{ChiralAni} gate is an extension of the Ani gate that additionally incorporates the full three-qubit chiral anisotropic interaction discussed in the next section, including $ZZZ$ components.  For gates implemented using the four-step protocol while utilizing the same interactions, we append \textbf{(Echo)} to the protocol name. We defer the discussion of the physical implementation of the interaction required for these gates to the next section, where we derive these interactions for spin qubit platforms.

\begin{figure}
    \centering
    \includegraphics[width= \linewidth]{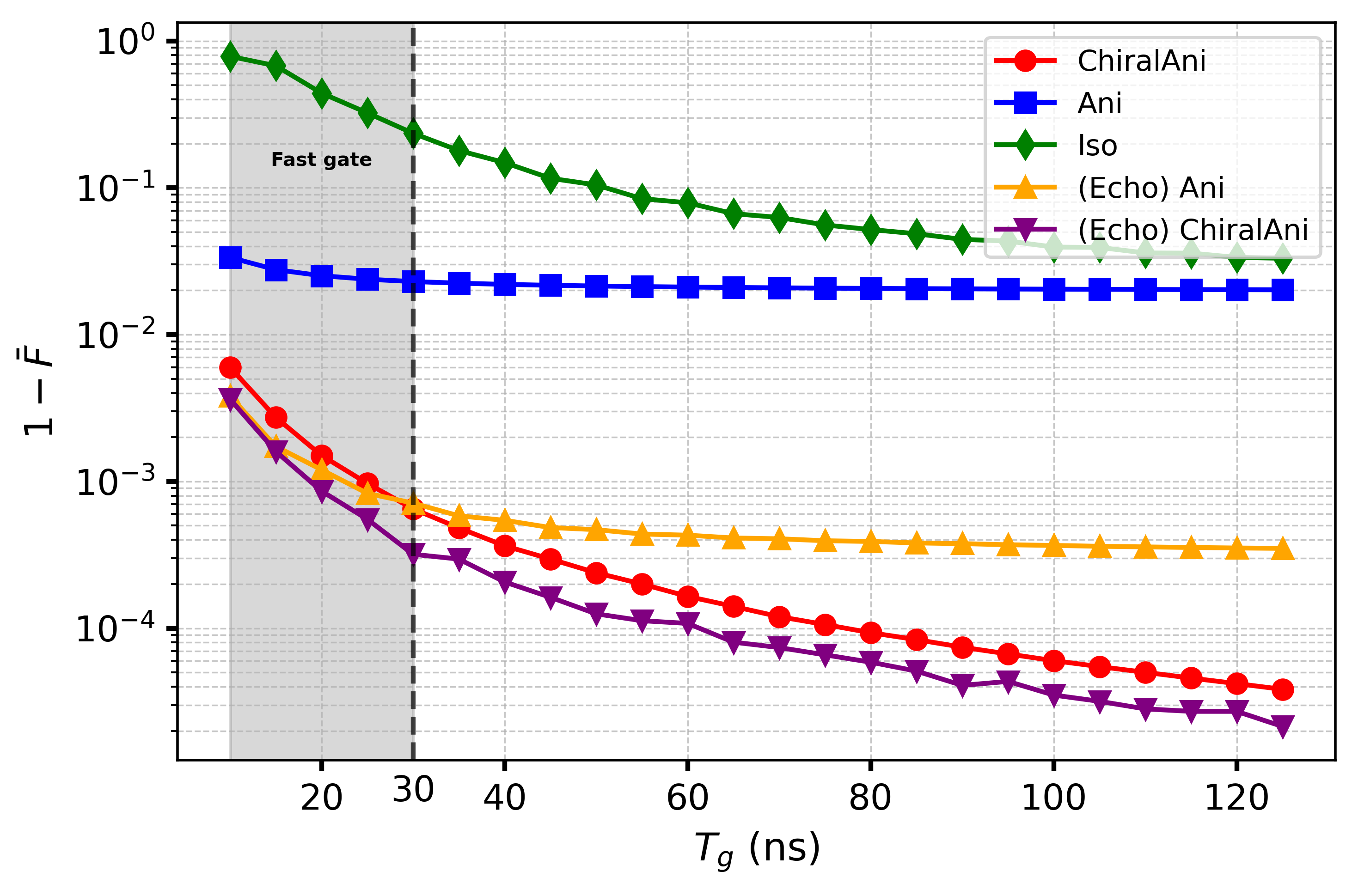}
    \caption{\justifying \textbf{Gate infidelity vs $T_g$ with flip-flop interactions. } We present the Iso gate (green), Ani gate (blue), ChiralAni gate (red), and their Echo variants (yellow and purple, respectively). We exclude qubit phase-correction gate times and single-qubit gate times for the Echo variants, as these are typically fast or can be accounted for with software. The plot is divided into two regions: (i) the fast regime (shaded), where the (Echo) Ani gate outperforms the ChiralAni gate, and (ii) the normal regime, where the ChiralAni gate performs better.}
    \label{fig: infidelity flip flop}
\end{figure}

For our numerical simulations, we set the qubit frequencies to $\omega_1/2\pi = 4.2~\rm GHz$,  $\omega_2/2\pi = 4.5~\rm GHz$, and  $\omega_3/2\pi = 4.8~\rm GHz$, which are typical values for spin-qubit devices~\cite{burkard2023semiconductor}. The ratios between the interaction strengths and Rabi frequencies are fixed through the $(0,1,1,2)$ synchronization condition in Eq.~\eqref{eq: synchronization ratios}. For the echo protocols, we employ a circuit-based approach by simulating each term in Eq.~\eqref{eq: Hahn Echo structure} separately and combining them at the end. This approach assumes the ability to rapidly switch the interactions on/off to implement the echo gate $Y_2$. We delay the relaxation of this assumption and a more detailed simulation, accounting for pulse timing errors, to future work.

\begin{table*}[htbp]
    \centering
    \renewcommand{\arraystretch}{1.1}
    \setlength{\tabcolsep}{6pt}
    \begin{tabular}{|c|c|c|c|c|c|c|}
        \hline
        $J_{12}$ & $J_{23}$ & $J_{13}$ & $J_{123}$ & Echo & $1-\bar{F}$ @ $30~\mathrm{ns}$ & $1-\bar{F}$ @ $100~\mathrm{ns}$ \\
        \hline
        Fi-Fo & Fi-Fo & N/A & N/A & $\times$ & $\approx 0.24$ & $\approx 0.04$ \\
        Fo-Fo & Fo-Fo & N/A & N/A & $\times$ & $\approx 0.023$ & $\approx 0.02$ \\
        Fo-Fo & Fo-Fo & N/A & N/A & $\checkmark$ & $\approx 7\times 10^{-4}$ & $\approx 4\times 10^{-4}$ \\
        Fo-Fo & Fo-Fo & Fi-Fo & ChiralAni & $\times$ & $\approx 7\times 10^{-4}$ & $\approx 6\times 10^{-5}$ \\
        Fo-Fo & Fo-Fo & Fi-Fo & ChiralAni & $\checkmark$ & $\approx 3\times 10^{-4}$ & $\approx 4\times 10^{-5}$ \\
        \hline
    \end{tabular}
    \caption{\justifying \textbf{Summary of gate protocol performance under flip-flop and flop-flop interactions.} Each protocol is defined by the off-diagonal elements of the pairwise exchange terms $J_{ij}$, either \textit{flip-flop} (Fi-Fo: $X_iX_j + Y_iY_j$) or \textit{flop-flop} (Fo-Fo: $X_iX_j - Y_iY_j$), as well as the presence of a chiral three-qubit interaction $J_{123}$ and the use of the proposed echo sequence. The final two columns report simulated average gate infidelity $1-\bar{F}$ at gate durations $T_g = 30~\mathrm{ns}$ and $T_g = 100~\mathrm{ns}$.}
    \label{tab: gate_infidelity_summary}
\end{table*}

In Fig.~\ref{fig: infidelity flip flop}, we plot the average fidelity of the $\rm C^2Ry(\pi)$ gate implemented with different protocols as a function of gate time $T_g$, which in turn determines the two-qubit exchange interaction $J_{\parallel}$ through Eq.~\eqref{eq: gate time with n}. The results demonstrate that the four-step Echo protocol significantly improves the gate fidelity for both the Ani gate and ChiralAni gate. Specifically, for the Ani gate, the Echo protocol reduces systematic errors by at least two orders of magnitude for moderately fast gate times ($T_g \geq 40~ \rm ns$). Additionally, the ChiralAni gate consistently outperforms both the Ani gate and its Echo variant, except in the fast regime for $T_g \leq 30~\rm ns$. However, the (Echo) ChiralAni gate provides the best fidelity across all gate times. From the plot, we also observe the impact of synchronization issues, where the fidelity of the Ani gate saturates around $T_g \sim 40~\rm ns$, while the ChiralAni gate continues to improve in fidelity with increasing gate time, achieving $\bar{F} \sim 0.9999$ with gate time $T_g \sim 80~\rm ns$, comparable to typical two-qubit gate times \cite{Gullans2019,Russ2018,Rasmussen2020}. 

In this subsection, we have numerically simulated the gate protocols and shown that the inclusion of a $ZZZ$ interaction reduces the gate time needed to achieve a target fidelity and improve the gate fidelity for a given gate time. For clarity, Table~\ref{tab: gate_infidelity_summary} summarizes the five gate protocols shown in Fig.~\ref{fig: infidelity flip flop}, including the required two- and three-qubit interactions and the average gate infidelities at $T_g = 30~\rm ns$ and $T_g = \rm 100~\rm ns$. However, our analysis has thus far only considered the systematic errors arising from flip-flop/flop-flop interactions present in the Hamiltonian. In Section~\ref{section: numerical results}, we extend this analysis by performing numerical benchmarking of the various gate protocols under more realistic physical noise models, namely $1/f$ noise  and calibration errors.

\section{Anisotropic chiral interaction}
\label{section: anisotropic chiral interaction}
\subsection{Physical model}
Many-body interactions have been proposed across various quantum platforms, including cold atom systems \cite{luo2024}, trapped ions \cite{Katz2023}, superconducting circuits \cite{Pederson2019}, and Andreev spin qubits \cite{lu2025kramersprotectedhardwareefficienterrorcorrection}. In this section, we demonstrate that a tunable three-qubit interaction emerges naturally in state-of-the-art QD spins with triangular dot arrangement, see Fig.~\ref{fig: triangular-device}. We show that  the $ZZZ$ interaction introduced in Sec.~\ref{section: ideal Hamiltonian} requires a non-trivial interplay of SOI and orbital magnetic field. Furthermore, we find that with these ingredients the three-qubit interaction naturally emerges and remains highly tunable over a broad range of Hamiltonian parameters. 

Our starting point is the generalized Fermi-Hubbard Hamiltonian 
\begin{subequations}
     \label{eq: Fermi-Hubbard Hamiltonian}
    \begin{align}
        H_{\rm FH}  & = \sum_{i,\sigma} \epsilon_i \hat{n}_{i,\sigma}-\sum_{i\neq j} \tau_{ij}  \textbf{c}_{i}^{\dagger} \hat{S}_{\rm rot}^{ij} \textbf{c}_j + \sum_{i} U_i  \hat{n}_{i,\up } \hat{n}_{i,\dw} \\
        & + \sum_{i}\frac{\omega_i}{2}(\hat{n}_{i,\up} - \hat{n}_{i,\dw})   ,
    \end{align}
    \begin{equation}
        \hat{S}_{\rm rot}^{ij}  = \exp\Big[- i \gamma_{ij}(\vec{n}_{\rm so, ij}\cdot \vec{\sigma}) \Big], 
    \end{equation}
\end{subequations}
describing a system with three tunnel-coupled QDs in the presence of SOI. Here, $\textbf{c}_{i} = [\hat{c}_{i,\up}, \hat{c}_{i,\dw}]^{T}$ denotes the spinor of $Q_i$ and $\vec{\sigma}$ is the vector of Pauli matrices. Each QD $Q_i$ has an energy $\epsilon_i$, a Zeeman energy splitting $\omega_i$, and spin-resolved density $\hat{n}_{i,\sigma}=\hat{c}_{i,\sigma}^\dagger \hat{c}_{i,\sigma}$. The tunneling between $Q_i$ and $Q_j$ has an energy $\tau_{ij}$ and also produces a spin rotation because of the SOI. This is described by the unitary operator $\hat{S}_{\rm rot}^{ij}$, that rotates the spin around the vector $\vec{n}_{\rm so,ij}$ by an angle $\gamma_{ij}$. This unitary satisfies the condition $\hat{S}_{\rm rot}^{ji} = (\hat{S}_{\rm rot}^{ij})^{\dagger}$ due to the hermicity of the Hamiltonian. 

Explicitly, we parametrize the SOI vector $\vec{n}_{\rm so,ij}$ as
\begin{equation}
    \vec{n}_{\rm so,ij} = \Big[ \sin(\theta_{ij} ) \cos(\varphi_{ij}), \sin(\theta_{ij} ) \sin(\varphi_{ij}), \cos(\theta_{ij}) \Big].
\end{equation}
For simplicity, in the following we assume that that the on-site potential $U_i = U$ is identical between all sites and consider the symmetric configuration with zero detuning $\epsilon_{i}=0$. This condition provides a sweet spot against detuning noise \cite{reed2016reduced,martins2016noise}, but is not required to our analysis and will be relaxed in Sec.~\ref{section: engineering anisotropic chiral interactions}.

We emphasize that the Fermi-Hubbard Hamiltonian in Eq.~\eqref{eq: Fermi-Hubbard Hamiltonian} is presented in the qubit-frame~\cite{Geyer2024}. Consequently, the tunneling matrix $\hat{S}^{ij}_{\rm rot}$ contains all the anisotropy in the system that arises from intrinsic SOI, tilted g-tensors, and local magnetic fields caused by micro and nano magnets.

\subsection{2-body anisotropic interaction}
Here, we briefly review the results of Ref.~\cite{Geyer2024} where the anisotropic SOI was used to perform fast and high fidelity two-qubit gates, and we adapt it to our three-spin setup. In the $(1,1,1)$ configuration and when $U \gg \tau$, an effective qubit Hamiltonian $H_{\rm eff}$ can be obtained perturbatively by using the Schrieffer–Wolff (SW) transformation \cite{tUexpansion}. This  Hamiltonian $H_{\rm eff}$ is decomposed into a series expansion
\begin{equation}
    \label{eq: t/U expansion}
    H_{\rm eff} = \sum_{i}\frac{\omega_i}{2} Z_i +   \sum_{n = 2}^{\infty} H_{\rm eff,n}
\end{equation}
where $||H_{\rm eff,n}||$ is of order $O\Big( \tau^{n}/U^{n-1}  \Big)$ and corresponds to the action of $n$ virtual tunneling events in the $(1,1,1)$ subspace. Without SOI (i.e $\gamma_{ij}=0$), we recover the usual Heisenberg exchange interaction 
\begin{equation}
     \label{eq: isotropic exchange interaction}
    H_{\rm eff,2}^{iso} = \sum_{\langle ij \rangle} \frac{J_{ij}}{4} \vec{\sigma}_j \cdot \vec{\sigma}_i = \sum_{ij} \frac{J_{ij}}{4} \begin{bmatrix}
        -1 & 0 & 0 & 0 \\
        0 & 1 & 2 & 0 \\
        0 & 2 & 1 & 0 \\
        0 & 0 & 0 & -1
    \end{bmatrix}_{ij},
\end{equation}
where the matrix representation is written in the subspace $\{ \ket{i=0/1,j=0/1} \} $ and $J_{ij} = 4 (\tau_{ij})^2/U$. The Heisenberg exchange interaction arises due to second-order virtual hopping processes, for example the off-diagonal elements in Eq.~\eqref{eq: isotropic exchange interaction} arise from the trajectories 
\begin{subequations}
    \begin{align}
    & \ket{\sigma \bar \sigma} \xrightarrow{} \ket{0,\sigma \bar\sigma} \xrightarrow{} \ket{\bar \sigma \sigma}. 
    \end{align} 
\end{subequations}
In general, the exchange interaction depends on the SOI vectors $\vec{n}_{so,ij}$ and is given by 
\begin{equation}
    H_{\rm eff,2}  = \sum_{\langle ij \rangle} \frac{J_{ij}}{4} \vec{\sigma}_j \cdot R(\gamma_{ij}, \theta_{ij},\varphi_{ij})\cdot \vec{\sigma}_i,
\end{equation}
where the explicit expression of the matrix $R(\gamma_{ij}, \theta_{ij},\varphi_{ij})$ is given in Eq.~\eqref{eq: full two-body exchange interaction} of Appendix.~\ref{appendix: miscroscopic model}. In Ref.~\cite{Geyer2024}, this anisotropic exchange interaction was utilized to enhance both the fidelity and speed of the controlled-rotation gates, compared to the conventional Heisenberg exchange interaction, by aligning the SOI vector $\vec{n}_{\rm so,ij}$ to lie in the equatorial plane of the Bloch sphere and operating at large $\gamma_{ij}\sim \pi/2$. This condition has also been proposed to improve baseband pulse control of the singlet-triplet qubit by suppressing leakage errors \cite{Spethman2024}. 

We now focus on this optimal condition and consider a fully anisotropic system where all SOI angles are $\gamma_{ij} = \pi/2$, and the SOI vectors $\vec{n}_{\rm so,ij}$ are perpendicular to the quantization axis. For such a system, the matrix representation of $H_{\rm eff,2}$ is given by
\begin{equation}
    \label{eq: anisotropic two-qubit exchange interaction}
    H_{\rm eff,2}^{ani} =\sum_{\langle ij \rangle}
    -\frac{J_{ij}}{4} \begin{bmatrix}
        1 & 0 & 0 & 2 e^{2 i \varphi_{ij} } \\
        0 & -1 & 0 & 0 \\
        0 & 0 & -1 & 0 \\
        2 e^{-2 i \varphi_{ijA} } & 0 & 0 & 1
    \end{bmatrix}.
\end{equation}
The flop-flop terms $\propto X_i X_j -Y_i Y_j$ (up to a single-qubit rotation around the $Z$-axis) couple the states $\{ \ket{0 0}_{ij}, \ket{1 1}_{ij}\}$ which have an energy gap $\omega_i + \omega_j$. As a result, the resonant gate condition requires the exchange interaction $J_{ij}$ to be small compared to $\omega_i + \omega_j$,  which is usually an order of magnitude larger than $|\omega_i - \omega_j|$ in the isotropic exchange interaction, enabling faster controlled-rotation gates. The smallness of $J_{ij}$ compared to the energy gap $\omega_i + \omega_j$ also ensures that the flop-flop terms in Eq.~\eqref{eq: anisotropic two-qubit exchange interaction} can be safely neglected when the gates are fast, justifying the ideal diagonal Hamiltonian introduced earlier in Eq.~\eqref{eq: ideal H two-body}. We also numerically verify this assumption in Fig.~\ref{fig: infidelity flip flop}, where the infidelity curve of the Ani gate remains relatively constant as we vary the strength of the anisotropic flop-flop term by changing the gate time $T_g$.

\subsection{3-body interaction}
\label{section: 3-body interaction}
\subsubsection{Isotropic chiral interaction}
We now extend our effective spin model beyond the effective two-qubit interactions by considering higher-order terms in the expansion of Eq.~\eqref{eq: t/U expansion}. 
In a triangular dot arrangement, the third-order expansion introduces an effective three-qubit interaction, which we show to be non-negligible, with a magnitude on the order of $\sim \rm MHz$. In contrast, higher-order contributions, such as the fourth-order effective Hamiltonian $H_{\rm eff,4}$, primarily introduce minor corrections (on the order of $\sim 1~\rm kHz$) to the two-body interactions \cite{mizel2004three} and are therefore neglected.

Without orbital magnetic field and phases $\Phi_B$, all odd-order qubit interactions vanish identically due to the chiral symmetry \cite{tUexpansion,Sen1995}.  An out-of-plane magnetic field $B^{\perp}_{\rm ext}$ applied to the device in Fig.~\ref{fig: triangular-device} breaks this symmetry and enables three-body interactions. The orbital magnetic field is introduced to the system through the Peierls phase  $\phi_{B}^{ij}$ of the tunneling coefficients  
\begin{equation}
    \tau_{ij} \to e^{i \phi_B^{ij}}~\tau_{ij} . 
\end{equation}
We note that the chiral symmetry can also be broken without $B^{\perp}_{\rm ext}$ via time-reversal symmetry-breaking Floquet driving \cite{sur2022driven,claassen2017dynamical,kitamura2017probing}. Adapting our three-qubit gate protocol to Floquet-driven systems is an interesting avenue for future research.

The effective three-qubit interaction arises due to the third-order virtual tunneling events, where the particle transverses the loop. In a system with no SOI $\gamma_{ij}=0$, the effective three-qubit interaction is the scalar chiral term \cite{Sen1995} 
\begin{equation}
    \label{eq: scalar chirality interaction}
    H_{\rm eff,3}^{iso} =  - \frac{3 \tau_{12} \tau_{23} \tau_{31}}{U^2}  \sin(\Phi_B)~\vec{\sigma}_1 \cdot (\vec{\sigma}_2 \times \vec{\sigma}_3),
\end{equation}
where $\Phi_B$ is the total magnetic field threading the triangular loop in Fig.~\ref{fig: triangular-device}. Physically, the factor $\sin(\Phi_B)$ arises from the Aharonov–Bohm interference between a particle’s trajectory and its time-reversed counterpart. For convenience, we define the three-qubit interaction strength in the absence of SOI as $J_{3Q}(\Phi_B)$, given by
\begin{equation}
    \label{eq: J3Q definition}
    J_{3Q}(\Phi_B) = \frac{24 \tau_{12} \tau_{23} \tau_{13}}{U^2} \sin(\Phi_B).
\end{equation}
With SOI, the three-qubit interaction strength is also proportional to  $ J_{3Q}(\Phi_B)$.

In typical experiments, the tunneling coefficients $\tau_{ij}$ can routinely reach $300 ~ \rm\mu eV$, and the on-site potentials $U$ range between $\sim 1-10~\rm meV$, resulting in two-qubit exchange interactions $J_{ij}/2\pi \sim 10-200~\rm MHz$ at zero detuning $\epsilon_i =0$. Likewise, after gauging out the modulation due to the orbital magnetic field, the three-qubit exchange interaction can reach values up to $J_{3Q}/2\pi \sim 1-30~\rm MHz$.

While the scalar chirality term has been proposed as a witness for tripartite entanglement \cite{Tsomokos2008} and as a mean to encode chiral qubits \cite{Scarola2004}, it has a negligible effect on the short timescale $T_{g}$ of the $\rm C^2 Ry $ gate. This is because the scalar chiral interaction not only has  an order of magnitude smaller amplitude than two-qubit interactions, but it is also completely off-diagonal and thus suppressed by the much larger differences of qubit energies. In the next section, we show that an effective SOI is essential to enable a diagonal $ZZZ$ component.

\subsubsection{Anisotropic chiral interaction with orbital magnetic field}
\label{section: chiral interaction with orbital magnetic field}

Before discussing in detail the derivation of the effective Hamiltonian $H_{\rm eff,3}$ with SOI, we provide a more intuitive physical explanation for the emergence of $ZZZ$ interactions.

Because of the fermionic anticommutation relations, our system presents two distinct destructive interference effects. The $ZZZ$ interaction emerges from the intricate interplay of both destructive interference effects.

The first comes from interference between a trajectory and its time-reversed version. Without an orbital magnetic field, chiral symmetry ensures $[\tau_{ij}]_{\sigma \sigma} = [\tau_{ji}]_{\bar \sigma \bar \sigma}$. With an orbital magnetic field, each trajectory accumulates a total magnetic phase $\Phi_{B}$ while traveling clockwise and $-\Phi_{B}$ counterclockwise. For example, the accumulated phases along a trajectory and its time-reversed counterpart are
\begin{subequations}
    \begin{align}
        & \ket{\up \dw \dw} \xrightarrow[e^{i \Phi_B/3}]{} \ket{0,\up \dw,\dw} \xrightarrow[e^{i \Phi_B/3}]{} \ket{0,\dw,\up \dw} \xrightarrow[e^{i \Phi_B/3}]{} \ket{\up \dw \dw},  \\
        & \ket{\up \dw \dw} \xrightarrow[e^{-i \Phi_B/3}]{} \ket{\up \dw,0, \dw} \xrightarrow[e^{-i \Phi_B/3}]{} \ket{\up \dw,\dw, 0} \xrightarrow[e^{-i \Phi_B/3}]{} \ket{\up \dw \dw}  .
    \end{align}
\end{subequations}
As a result, the longitudinal interaction $J_{123}^{\parallel}$ acquires the factor $\sin(\Phi_B)$ because of  the Aharonov-Bohm effect. 

The second interference effect arises from the spin-dependent phase difference induced by SOI. For simplicity, we assume that all SOI vectors $\vec{n}_{\rm so,ij}$ point in the $Z$-direction. In this case, as a spin tunnels from one dot to the next, it accumulates a certain spin-dependent phase because of SOI, in particular a spin-up (spin-down) particle accumulates a phase $e^{-i \gamma_{ij}}$ ($e^{i\gamma_{ij}}$) while tunneling from $Q_i$ to $Q_j$. This phase accumulation has an effect when the spin completes a closed loop around the triangular arrangement of QDs. An example of two closed trajectories with the same chirality (i.e., the same Aharonov-Bohm phase) is
\begin{subequations}
    \begin{align}
        & \ket{\up \dw \dw} \xrightarrow[e^{-i\gamma_{12}}]{} \ket{0,\up \dw,\dw} \xrightarrow[e^{-i\gamma_{23}}]{} \ket{0,\dw,\up \dw} \xrightarrow[e^{-i\gamma_{13}}]{} \ket{\up \dw \dw},  \\
        & \ket{\up \dw \dw} \xrightarrow[e^{i\gamma_{13}}]{} \ket{\up \dw, \dw,0} \xrightarrow[e^{i\gamma_{23}}]{} \ket{\up \dw,0, \dw} \xrightarrow[e^{i\gamma_{12}}]{} \ket{\up \dw \dw} .
    \end{align}
\end{subequations}
Consequently, the longitudinal interaction $J_{123}^{\parallel}$ acquires a factor of $\sin(\sum_{\langle ij \rangle} \gamma_{ij})$ due to the SOI-induced phase difference between spin-up and spin-down particles. In Appendix~\ref{appendix: proof of infeasiblity}, we explicitly derive the dependence of the longitudinal interaction $J_{123}^{\parallel}$ on third-order virtual tunneling amplitudes, see e.g. Eq.~\eqref{eq: amplitude J123 from tunneling events}.

We now validate this intuitive picture by a detailed derivation based on the Fermi-Hubbard model in Eq.~\eqref{eq: Fermi-Hubbard Hamiltonian}. In this subsection, we simplify our discussion by assuming that the magnetic field only produces an orbital contribution and does not lead to the Zeeman energy splitting.  This allows us to clearly illustrate the underlying physical mechanism and the emergence of the anisotropic exchange interaction. In the next subsection, we extend our discussion to include the effects of Zeeman splitting and arbitrary SOI configurations.

As the first step, we perform a set of local rotations $\hat{\mathcal{R}}_{i}$ of the quantization axis of each qubit $Q_i$ to move to a frame where all the SOI vectors $n_{\rm{so},ij}$ are aligned to the positive $Z$-direction and the SOI-rotation angles between any ordered pair of qubits are identical, i.e.
\begin{equation}
    \label{eq: SOI under basis transformation}
    \hat{\mathcal{R}}_{j}^{\dagger} \hat{S}^{ij}_{\rm rot} \hat{\mathcal{R}}_{i} = \text{exp}\left({-i\tilde{\gamma}\sigma_z}\right) \equiv \hat{S}_{\rm ave}\ .
\end{equation}
We call this the Easy Spin-Orbit Axis (ESOA) frame, and we introduce the average SOI spin rotation $ \hat{S}_{\rm ave}$ and angle $\tilde \gamma$. 
The existence of the transformation is proved in Appendix.~\ref{appendix: miscroscopic model}.  Because of the closed-loop connectivity, $\hat{S}_{\rm ave}$ is unitarily equivalent to the geometric mean of the set $\{ \hat{S}^{ij}_{\rm rot} \}$ 
\begin{equation}
    \label{eq: cyclic-condition of ESOA frame}
    \hat{S}_{\rm ave}^{3} = \hat{\mathcal{R}}_{1}^{\dagger}  \hat{S}^{31}_{\rm rot}   \hat{S}^{23}_{\rm rot}   \hat{S}^{12}_{\rm rot} \hat{\mathcal{R}}_{ 1} ;
\end{equation}
similar equations can be derived for the other rotations $\hat{\mathcal{R}}_{i}$ by cyclic permutations. In general, we can efficiently compute the average angle $\tilde\gamma$ and the local rotations as the eigensystems of the cyclic permutations of Eq.~\eqref{eq: cyclic-condition of ESOA frame}. We also remark that the existence of a non-trivial ESOA frame is strongly dependent on the closed-loop connectivity: for linear connectivity a similar spin-orbit transformation results in $\hat{S}_{ave} = I$ \cite{Geyer2024,bosco2024exchangeonlyspinorbitqubitssilicon,rimbachruss2024spinlessspinqubit}. 

From the operator $\hat{S}_{\rm ave}$, we obtain the effective Hamiltonian $\tilde{H}_{\rm eff,3}$ in the ESOA frame by performing a third-order SW transformation on the Fermi-Hubbard Hamiltonian in Eq.~\eqref{eq: Fermi-Hubbard Hamiltonian}, leading to
% Backhand calculation shows that roughly 50 mT is enough
\begin{subequations}
    \label{eq: exact chiral anisotropic interaction}
    \begin{align}
    & \tilde{H}_{\rm eff,3} = -\frac{\tau_{12} \tau_{23} \tau_{13}}{U^2} \sin(\Phi_B) \Big[ -\sin(3 \tilde{\gamma})\sum_{i=1}^{3} Z_i  \\
    \label{eq:  exact chiral anisotropic interaction off-diagonal}
    &  +  \sum_{\langle ijk \rangle }  Z_i \begin{bmatrix}  X_j \\ Y_j \\ Z_j \end{bmatrix}^{T}\begin{bmatrix}   3 \sin(\tilde \gamma ) & 3 \cos(\tilde \gamma) &0 \\    -3 \cos(\tilde \gamma) & 3  \sin(\tilde \gamma) & 0 \\ 0 & 0 & \sin(3 \bar{\gamma}) \end{bmatrix}\begin{bmatrix}   X_k \\ Y_k \\ Z_k \end{bmatrix} \Big].
\end{align}
\end{subequations}
We notice that the $ZZZ$ interaction is preceded by the prefactor $\sin(3\tilde{\gamma})$, which confirms that this interaction is indeed due to the phase-interference effect between a spin-up particle and a spin-down particle after traversing the triangular loop. We also find that the effective Hamiltonian $\tilde{H}_{\rm eff,3}$ contains the factor $\sin(\Phi_B)$ due to chirality symmetry breaking. 
The effective Hamiltonian $\tilde{H}_{\rm eff,3}$ also contains single-qubit $Z$ operators and a modified scalar chirality term. The combination of the Pauli-$Z$ operators and the $ZZZ$ term ensures that there is no third-order virtual tunneling events if the spin-configuration is $\ket{s,s,s}$. Physically, the Pauli-$Z$ operators produce small deviations to the qubit frequencies, which can be compensated for, and will be neglected in the following discussion. We also safely neglect the modified scalar chirality term (i.e in the upper-left block of Eq.~\eqref{eq:  exact chiral anisotropic interaction off-diagonal}) that is fully off-diagonal, and strongly suppressed by the large Zeeman energy difference, as we argued in Sec.~\ref{section: ideal Hamiltonian}.

We then approximate $\tilde{H}_{\rm eff,3}$ during the $\rm C^2 Ry$ gate duration $T_{g}$ as 
\begin{equation}
    \label{eq: effective H2 in ESA frame}
    \tilde{H}_{\rm eff,3} \approx - \frac{J_{3Q}(\Phi_B)}{8} \sin(3 \tilde{\gamma}) Z_1 Z_2 Z_3.
\end{equation}
where $J_{3Q}(\Phi_B)$ is the amplitude of the three-body interaction previously defined in Eq.~\eqref{eq: J3Q definition}. 

In this subsection, we explained the physical mechanism of the anisotropic exchange interaction and demonstrated that it can be tuned by engineering the SOI. In the next subsection, we show that our analysis naturally extends to scenarios where we do not have SOI but the qubit quantization axes are different. 
\subsubsection{Anisotropic exchange interaction with tilted g-tensor}
\label{section: chiral interaction with tilted g-tensor}
In the previous section, we assumed that the qubit quantization axes are identical and that the SOI can be controlled to realize anisotropic exchange interactions. Here, we explore the opposite regime, where individual qubit $g$-tensors are tilted and distinct, but SOI is absent. This scenario is particularly relevant to recent experiments \cite{Chien-An2024, Rooney2025, Zhang2025, saezmollejo2024exchangeanisotropiesmicrowavedrivensinglettriplet} in germanium spin-qubit platforms as well as silicon spin-qubits with micromagnets and low magnetic fields \cite{unseld2024baseband, philips2022universal}, which have demonstrated precise measurements of the anisotropic $g$-tensors and control over local magnetic fields, respectively. 

In this case, the lab-frame Hamiltonian up to a third-order SW transformation and assuming no spin-flip tunneling, is given by 
\begin{equation}
    \label{eq: lab frame tilted g tensor}
    H_{\rm lab} = \sum_{i=1,2,3} \frac{\mu_B \vec{B}_i \cdot \hat{g}_i }{2}\cdot \vec{\sigma}_i + \sum_{ \langle ij \rangle} \frac{  J_{ij}}{4} \vec{\sigma}_i\cdot\vec{\sigma}_j  - \frac{J_{123}}{8} \vec{\sigma}_1\cdot(\vec{\sigma}_2 \times \vec{\sigma}_3),
\end{equation}
where $\hat{g}_i$ are the qubit $g$-tensor, $\vec{B}_i$ are the local magnetic fields, and $\mu_B$ is the Bohr magneton. To transform the lab frame into the qubit frame—where the single-qubit terms reduce to simple Zeeman energy splittings—we apply a set of local spin-Pauli rotation matrices $\{ R_{Q,i} \}$ that depend on the local Zeeman fields $\propto \vec{B}_i \cdot \hat{g}_i$ \cite{Geyer2024,bosco2024exchangeonlyspinorbitqubitssilicon,rimbachruss2024spinlessspinqubit}. The qubit frequencies in the qubit frame are given by 
\begin{equation}
    \omega_i = | \mu_B \vec{B}_i\cdot\hat{g}_i\cdot R_{Q,i}| \ .
\end{equation}

 An effective SOI emerges in the qubit frame due to spin-flip processes induced by the misalignment of quantization axes between the QDs [see also Eq.~\eqref{eq: SOI under basis transformation}]. In the qubit frame, the scalar chirality interaction in  Eq.~\eqref{eq: lab frame tilted g tensor} is transformed as 
\begin{equation}
    \vec{\sigma}_1\cdot(\vec{\sigma}_2 \times \vec{\sigma}_3) \to    (R_{Q,1 }\vec{\sigma}_1)\cdot \Big[ (R_{Q,2 }\vec{\sigma}_2) \times (R_{Q,3} \vec{\sigma}_3) \Big],
\end{equation}
which generally includes a $ZZZ$ interaction when all the rotation matrices $\{ R_{Q,i} \}$ are distinct. 

We emphasize that if any of the rotation matrices $\{ R_{Q,i} \}$ are identical, the system does not support the $ZZZ$ interaction. To understand this condition, we consider $R_{Q,1} = R_{Q,3}$. A key consequence is that the SOI operator $\hat{S}_{rot}^{12}$ between qubits $Q_1 \leftrightarrow Q_2$ is the conjugate-transpose of the SOI operator $\hat{S}_{rot}^{23}$ between qubits $Q_2 \leftrightarrow Q_3$, since 
\begin{equation}
    R_{Q,2}^{T} R_{Q,1} = (R_{Q,3}^{T} R_{Q,2})^{T}.
\end{equation}
Moreover, there is no effective SOI between qubits $Q_1 \leftrightarrow Q_3$. As a result, when a particle tunnels from $Q_2$ to $Q_3$, it acquires a phase $\gamma_{23} = - \gamma_{12}$, opposite to the phase obtained when tunneling from $Q_{1}$ to $Q_{2}$. Meanwhile, no spin-dependent phase is acquired when traveling from  $Q_3$ to $Q_1$. More explicitly, we can consider the tunneling trajectory
\begin{equation}
   \ket{\up \dw \dw} \xrightarrow[e^{-i\gamma_{12}}]{} \ket{0,\up \dw,\dw} \xrightarrow[ e^{i\gamma_{12}}]{} \ket{0,\dw,\up \dw} \xrightarrow[1]{} \ket{\up \dw \dw}
\end{equation}
and observe that the total accumulated phase vanishes; hence, no $ZZZ$ interaction emerge.

\subsection{Engineering anisotropic chiral interactions}
\label{section: engineering anisotropic chiral interactions}
\begin{figure}
    \centering
    % subfigure (a)
    % Subfigure (b)
    \begin{subfigure}{\linewidth}
        \centering
        \subcaption{\centering $J_{123}^{\parallel}(\gamma,\theta)/J_{3Q}(\Phi_B)$ }
          \includegraphics[width=0.95 \linewidth]{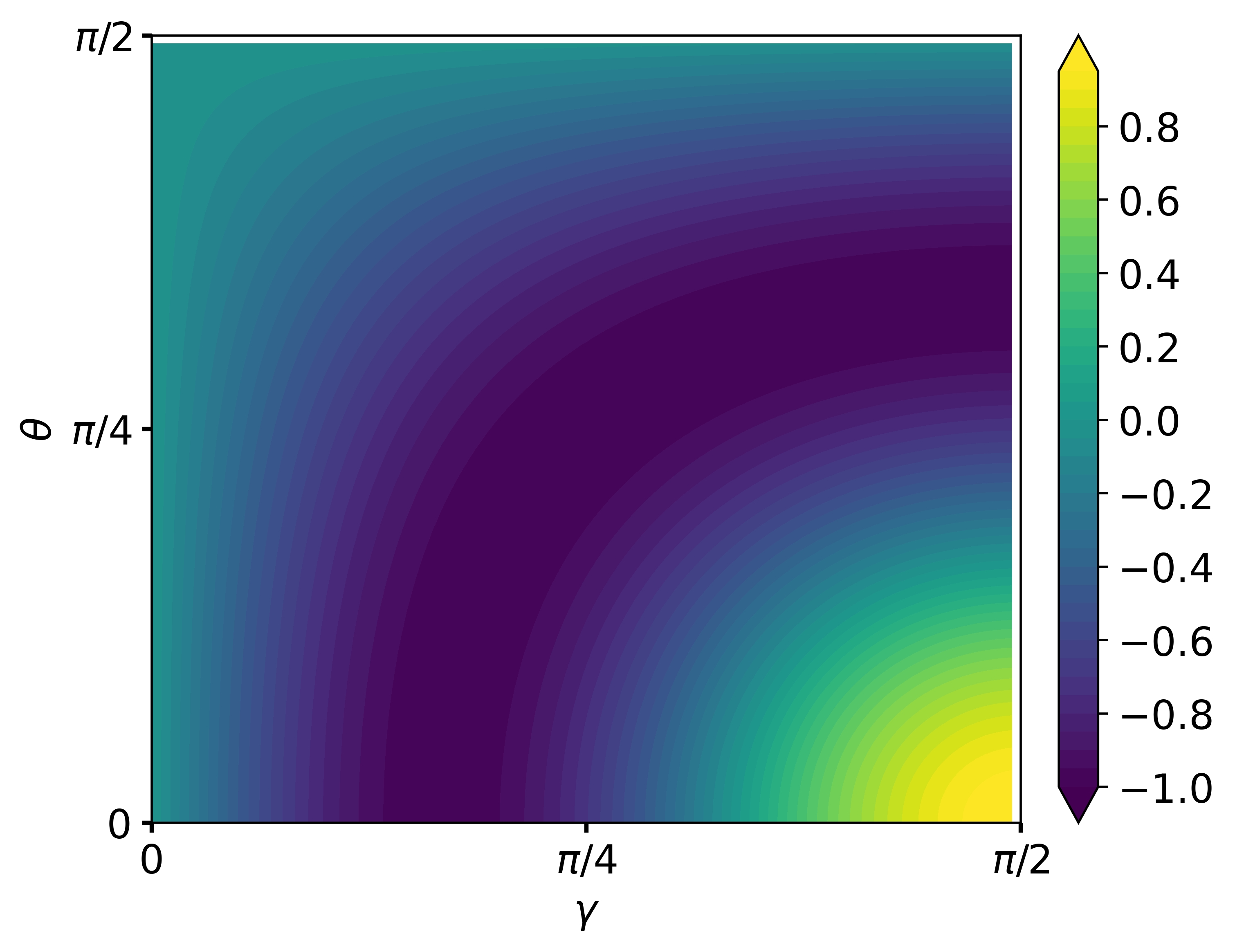}
        \label{fig: contour_SOI_angle_plot}
    \end{subfigure}
    \hfill
     % Subfigure (c)
    \begin{subfigure}{\linewidth}
        \centering
        \subcaption{\centering $J_{123}^{\parallel}(\epsilon_1,\epsilon_3)/J_{123}^{\parallel}(0,0)$}
        \includegraphics[width=0.95\linewidth]{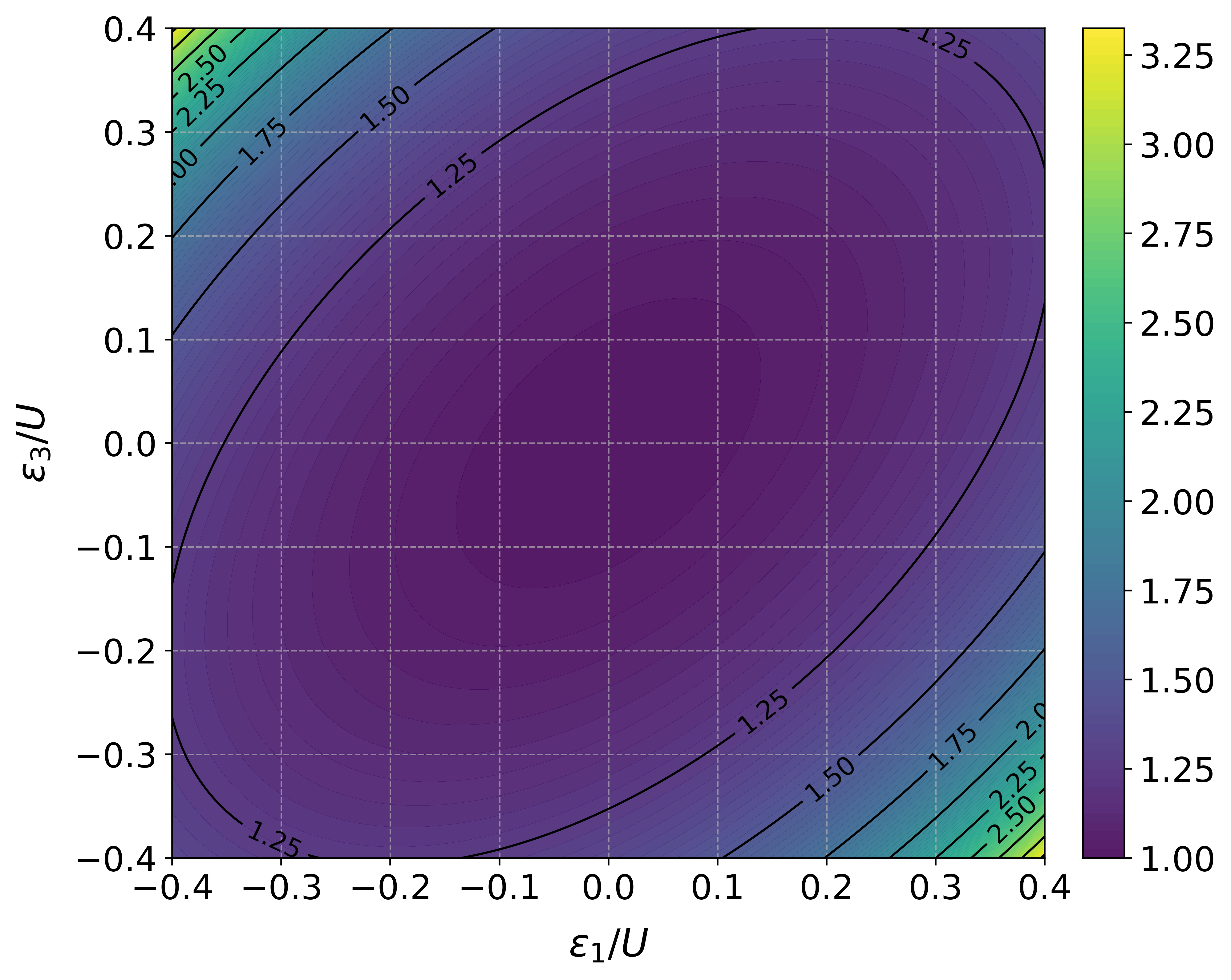}
        \label{fig: contour_detuning_plot}
    \end{subfigure}%
    \hfill
        \begin{subfigure}{\linewidth}
        \centering
        \subcaption{\centering Optimal SOI for $\rm C^2 Ry$ gates.}
        \includegraphics[width=0.95\linewidth]{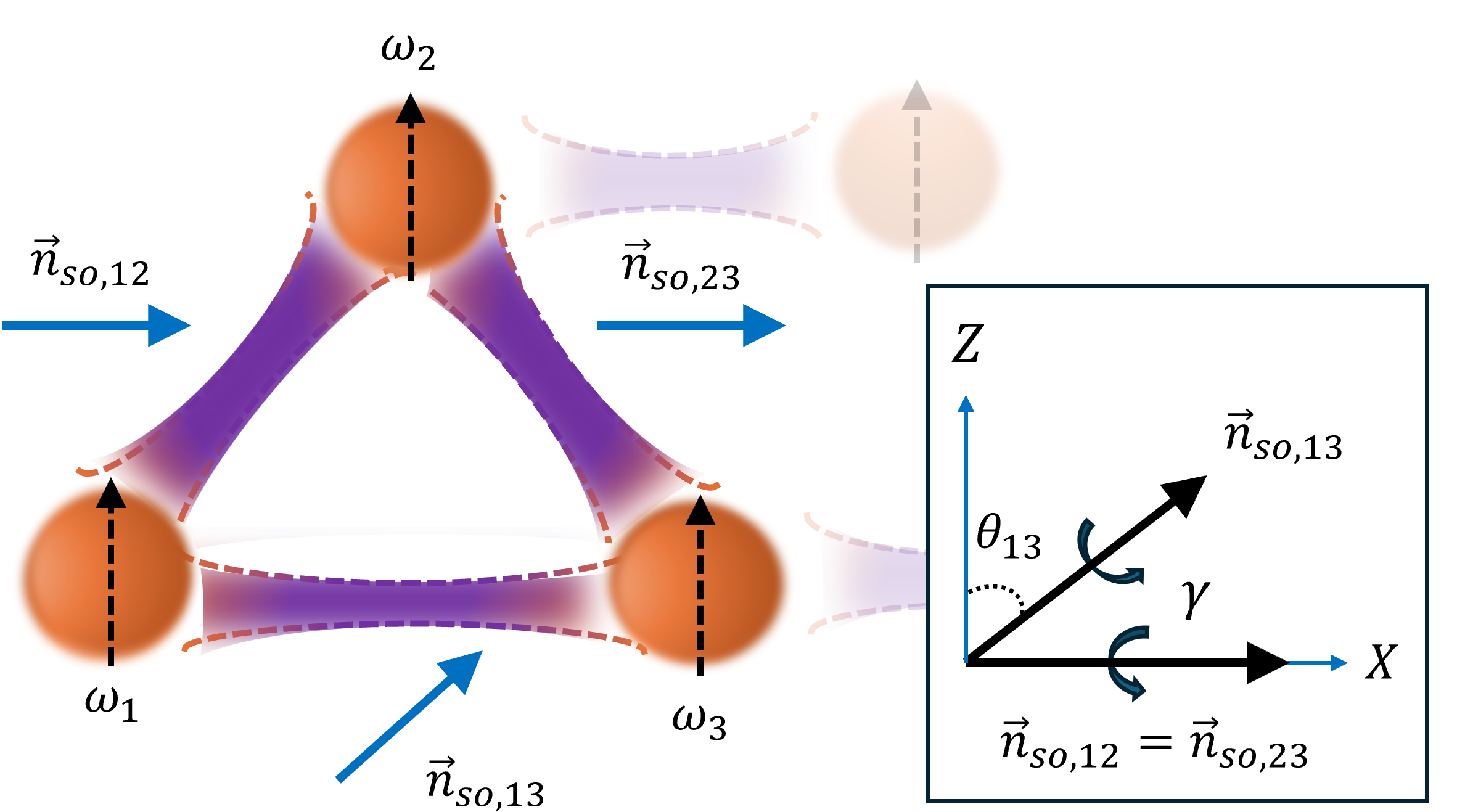}
        \label{fig: SOI configuration plot}
    \end{subfigure}
    \caption{\justifying \textbf{Dependence of $J_{123}^{\parallel}$ on SOI and QD energy detuning} (a) The ratio $J_{123}^{\parallel}/J_{3Q}(\Phi_B)$ for a symmetric system with $(\gamma_{ij},\theta_{ij},\varphi_{ij}) =(\gamma,\theta,\varphi)$. This ratio is independent of $\varphi$. (b) The ratio  $ J_{123}^{\parallel}(\varepsilon_{1},\varepsilon_{3})/J_{123}^{\parallel}(0,0)$, where $\varepsilon_{i}$ are the dot energies. This ratio is independent of the SOI configuration. (c) Sketch of the SOI configuration considered in Eq.~\eqref{eq: J123 for optimal SOI}. The optimal configuration for the  $\rm C^2Ry(\pi)$ gate is obtained at $\theta_{13}=0$ and $\gamma = \pi/2$.}
    \label{fig: contour_plot}
\end{figure}

\subsubsection{Tunable $ZZZ$ interaction}
In this section, we combine Zeeman energy and SOI, and present ways to engineer the $ZZZ$ interaction in state-of-the-art devices. We move again to the qubit frame, where the quantization axes align with the Zeeman field. In this frame, the SOI between two different qubits are
\begin{equation}
    S_{Q,j}^{\dagger}\hat{S}_{\rm rot}^{ij} S_{Q,i} \ ,
\end{equation}
where $S_{Q,i}$ is the unitary operator that generates the rotation matrix $R_{Q,i}$, i.e. $S_{Q,i}^\dagger \vec{\sigma}_i S_{Q,i}= R_{Q,i}\cdot\vec{\sigma}_i$.  Due to the large frequency differences between qubits, we focus primarily on the $ZZZ$ component, in analogy to the $ZZ$ component in the anisotropic 2-body interaction and neglect the flip-flop terms. This approach eliminates the need for fine-tuning to achieve only $ZZZ$ interaction.

The most efficient way to obtain a large $ZZZ$ interaction in the qubit frame is to align all spin-orbit vectors $\vec{n}_{\rm so,ij}$  in the $Z$ direction. However, this is not the only possibility and, as we will show, it is not the optimal choice for three-qubit gates. As we discussed in Sec.~\ref{section: chiral interaction with orbital magnetic field} and \ref{section: chiral interaction with tilted g-tensor}, chiral anisotropic interactions naturally emerge in systems such as hole gases that present large intrinsic SOI and anisotropic $g$-tensors, and in electronic systems with magnets.

In Fig.~\ref{fig: contour_SOI_angle_plot}, we show how  $J_{123}^{\parallel}$ depends on the SOI vectors and angles when the system is symmetric and $\gamma_{ij} = \gamma, ~\theta_{ij} = \theta, ~\varphi_{ij} = \varphi$. As discussed previously, the three-body interaction could reach  $J_{3Q}/2\pi \sim 1-30~\rm MHz$ in current experiments. A general triangular QD arrangement present sizable $ZZZ$ interactions for a wide range of parameters, as demonstrated by the large dark region in the figure. These interactions only vanish in some fine-tuned scenarios, which include the trivial case with no SOI and aligned Zeeman fields $(\gamma =0)$, and the case where all the Zeeman fields are aligned and perpendicular to the SOI vectors $(\theta=\pi/2)$.

We further highlight the tunability of the chiral anisotropic exchange interaction by analyzing numerically its dependence on the detuning between different QDs. In Fig.~\ref{fig: contour_detuning_plot}, we show the relative change in the magnitude of $J_{123}^{\parallel}$ as a function of two different QD energies  $\epsilon_{1}$ and  $\epsilon_{3}$ when  $\epsilon_{2}=0$, which is analytically given by 
\begin{equation}
    \frac{J_{123}^{\parallel}(\epsilon_1,\epsilon_3)}{J_{123}^{\parallel}(0,0) } = \frac{U^4(3U^2 - \epsilon_1^2+\epsilon_1 \epsilon_3-\epsilon_3^2)}{3(U^2-\epsilon_1^2) (U^2-\epsilon_3^2)[U^2-(\epsilon_1-\epsilon_3)^2]  }. 
\end{equation}
 % \ste{the explicit expression $J_{123}^{\parallel}(\epsilon_{1},\epsilon_{3})/J_{123}^{\parallel}(0,0)$ you can show directly in the figure }
This figure reveals two key insights. First, $J_{123}^{\parallel}$ remains stable up to $|\epsilon_{i}/U| \leq 0.1$, demonstrating the robustness of $ZZZ$ interactions against systematic detuning errors. Second,  the magnitude of $J_{123}^{\parallel}$ is highly tunable not only by modulating the tunnelings $\tau_{ij}$, but also by modifying local QD energies $\epsilon_i$. In particular, we observe an enhancement by more than a factor of two even with moderate detuning well within the (1,1,1) charge configuration. However, we note that  away from the center of the stability diagram, we are also moving away from the charge noise sweet spot.

Additionally, we emphasize that the chiral anisotropic interaction can be  turned on and off on-demand. For example, by electrically deactivating any of the tunnelings $\tau_{ij}$ \cite{ivlev2025operating,martins2016noise,reed2016reduced}, we can efficiently break the loop and prevent the third-order virtual tunneling process.  Alternatively, the chiral interaction can be suppressed by aligning the magnetic field in-plane, thereby restoring chiral symmetry. However, dynamically switching the magnetic field orientation from out-of-plane to in-plane is experimentally challenging.

\subsubsection{Optimal setup}

The symmetric configuration with aligned SOI vectors is useful to understand the key handles that determine the amplitude of the $ZZZ$ interaction. However, this is not the only possibility to reach large values of this  interaction and it is not the optimal choice for the three-qubit gates.
In particular, recent advances in $g$-tensor and SOI engineering \cite{BoscoFinfet,bosco2022fully,liu2022gate,malkoc2022charge,abadillo2023hole,secchi2021inter,bosco2021squeezed,terrazos2021theory,Chien-An2024,martinez2022hole,bosco2021fully,jirovec2022dynamics} and in nanomagnet design \cite{aldeghi2023modular,unseld2024baseband} suggest that additional fine-tuning of many-body interactions can become experimentally feasible \cite{Geyer2024,Geyer2021,Xue2022,Chien-An2024,Rooney2025,Zhang2025,saezmollejo2024exchangeanisotropiesmicrowavedrivensinglettriplet,carballido2024compromise,bassi2024optimal,liles2021electrical,Liles2024,voisin2016electrical}. 

With this in mind, we now propose a optimal setup for the  $\rm C^2Ry(\pi)$ gate that minimizes two-qubit flip-flop terms while maximizing the strength of the $ZZZ$ interaction. This optimal setup is sketched in  Fig.~\ref{fig: SOI configuration plot}. To achieve a fast gate time, we align the SOI vectors $\vec{n}_{so,12}=\vec{n}_{so,23}$ to lie in the equatorial plane, suppressing flip-flop interactions between $Q_1 \leftrightarrow Q_2$ and $Q_2 \leftrightarrow Q_3$, and both SOI vectors points in the same direction. The other vector $\vec{n}_{so,31}$ is out of plane.

In this case, assuming for simplicity identical SOI rotation angles, $\gamma_{ij} = \gamma$, the $ZZZ$ interaction strength is given by 
\begin{equation}
    \label{eq: J123 for optimal SOI}
    J_{123}^{\parallel} = -J_{3Q}(\Phi_B) \cos(\theta_{13}) \sin(\gamma).
\end{equation}
The optimal configuration to maximize interaction strength $J_{123}^{\parallel}$ 
while maintaining high gate fidelity is to align $\vec{n}_{so,13} $ parallel to the $Z$ axis ($\theta_{13} =0$) and adjust the SOI rotation angle to approximately $\gamma \sim \pi/2$.

Assuming a center-to-center QD distance of approximately $d_{QD} =100~\rm nm$ \cite{Acuna3holeQ,Chien-An2024}, the required (SOI) length $\lambda_{SO}$ to achieve an SOI rotation angle of $\gamma \approx d_{QD}/\lambda_{SO}\sim \pi/2$ is given by $\lambda_{SO} = 2d_{QD}/\pi \sim 63~\rm nm$. This value of SOI length $\lambda_{SO}$ falls within the range of previously reported experimental measurements \cite{Geyer2024,Camenzind2022,Geyer2021,Li2015}.

Focusing on this optimal SOI configuration for the $\rm C^2 Ry(\pi)$ gate, we can also estimate the required orbital magnetic field $B^{\perp}_{ext}$. The synchronization ratios from Eq.~\eqref{eq: synchronization ratios} impose the following constraint:
\begin{equation}
    \label{eq: ratio between J12 and J123}
    \frac{J_{12}^{\parallel}}{J_{3Q}(\Phi_B)} = \frac{U}{6 \tau_{13} \sin(\Phi_B)} \approx 20.
\end{equation}
To suppress unwanted flip-flop terms between $Q_1$ and $Q_3$ caused by the exchange interaction $J_{13}$, we restrict $\tau_{13}/U$  to lie within the range $[0.02,0.05]$. Given our previous assumption that the distance between the QDs is $d_{QD} = 100~\rm nm$, we estimate the required orbital magnetic field $B^{\perp}_{ext}$ to be in the range of $20~\rm mT$ to $60~\rm mT$. We also note that the required orbital magnetic field can be reduced by considering an alternative synchronization condition $(0,n,n,2n)$ with $n \geq 2$, see Eq.~\eqref{eq: (0nn2n) condition}. The resulting gate will have lower gate speed but higher gate-fidelity due to the flip-flop terms scale inversely with $n$, see also Figs.~\ref{fig: ideal gate infidelity} and \ref{fig: infidelity flip flop}. In the following, we show that the required orbital magnetic field is compatible with current silicon and germanium spin qubit devices.

In silicon qubits, the $g$-tensor is mostly isotropic with $g \approx 2$ and minimal variations between the QDs. If we assume an external magnetic field $\vec{B}_{\rm ext} \sim 300~\rm mT$ \cite{unseld2024baseband,Xue2022,philips2022universal,undseth2023nonlinear,takeda2022quantum,Geyer2024}, tilted $4^{\circ}$ \cite{Chien-An2024,john2024two} out of the $xy$-plane, we obtain $B^{\perp}_{\rm ext}\sim 21~\rm mT$, which falls within the required range for the synchronization ratio in Eq.\eqref{eq: ratio between J12 and J123}. At this field strength, the average qubit frequency is $\omega_i/2\pi\sim 8.6~\rm GHz$ and local field gradients can lead to frequency differences up to $200-300~\rm MHz$ \cite{unseld2024baseband,Xue2022,philips2022universal,undseth2023nonlinear,takeda2022quantum,Geyer2024}. While the average qubit frequency differs from our simulations (Fig.~\ref{fig: infidelity flip flop}), the qubit frequency differences are of the same order. This is the key factor to suppress flip-flop interactions.

In planar germanium qubits, the $g$-tensor is highly anisotropic, with an in-plane g-factor $g_{xy}\sim 0.4$ and an out-of-plane g-factor $g_{z}\sim 10$, varying between QDs by $\delta g_{xy} \sim 0.1$ and $\delta g_{z} \sim 0.5$ \cite{Zhang2025,hendrickx2024sweet,hendrickx2021four,Chien-An2024,john2024two,saezmollejo2024exchangeanisotropiesmicrowavedrivensinglettriplet,Rooney2025}. We consider two scenarios: (i) an out-of-plane magnetic field $B^{\perp}_{\rm ext} \sim 30~\rm mT$ yields an average qubit frequency of $\sim 4.2~\rm GHz$ and a qubit frequency difference of $\sim 170~\rm MHz$ due to $\delta g_{z}$; (ii) a mostly in-plane field $B_{\rm ext} \sim 300~\rm mT$ tilted by $\sim 4^{\circ}$ out-of-plane shifts the quantization axis to $\sim 53.7^{\circ}$ out-of-plane and results in an average qubit frequency of $\sim 3.3~\rm GHz$ with qubit frequency differences $\sim 200~\rm MHz$ caused by both $\delta g_{xy}$ and $\delta g_{z}$. These parameters match well those used in our simulations.

We also remark that precise control of inter-dot couplings in germanium bilayer devices \cite{ivlev2024coupled} can provide a natural way to reach sizable chiral anisotropic interactions as in-plane magnetic fields $B^{\parallel}_{\rm ext}$ thread the loop automatically.
 
In summary, in this section, we demonstrated that the combination of anisotropic interactions and orbital magnetic fields can give rise to chiral anisotropic interactions, which naturally emerge in state-of-the-art QD devices. We discussed how the chiral anisotropic interaction includes a $ZZZ$ interaction, enabling the realization of fast and high-fidelity $\rm C^2 Ry$ gate. In the next subsection, we present a protocol for experimentally measuring the three-qubit interaction strength $J_{123}^{\parallel}$ with high accuracy that is applicable to any generic three-qubit Hamiltonian.

\subsection{Measurement of the three-qubit interaction}
\begin{figure}
    \centering
    % Subfigure (a)
    \begin{subfigure}{\linewidth}
        \centering
        \subcaption{\centering Exchange oscillation $P_{DD}(\ket{+00},T_Z)$ from $J_{123}^{\parallel}$}
          \includegraphics[width=\linewidth]{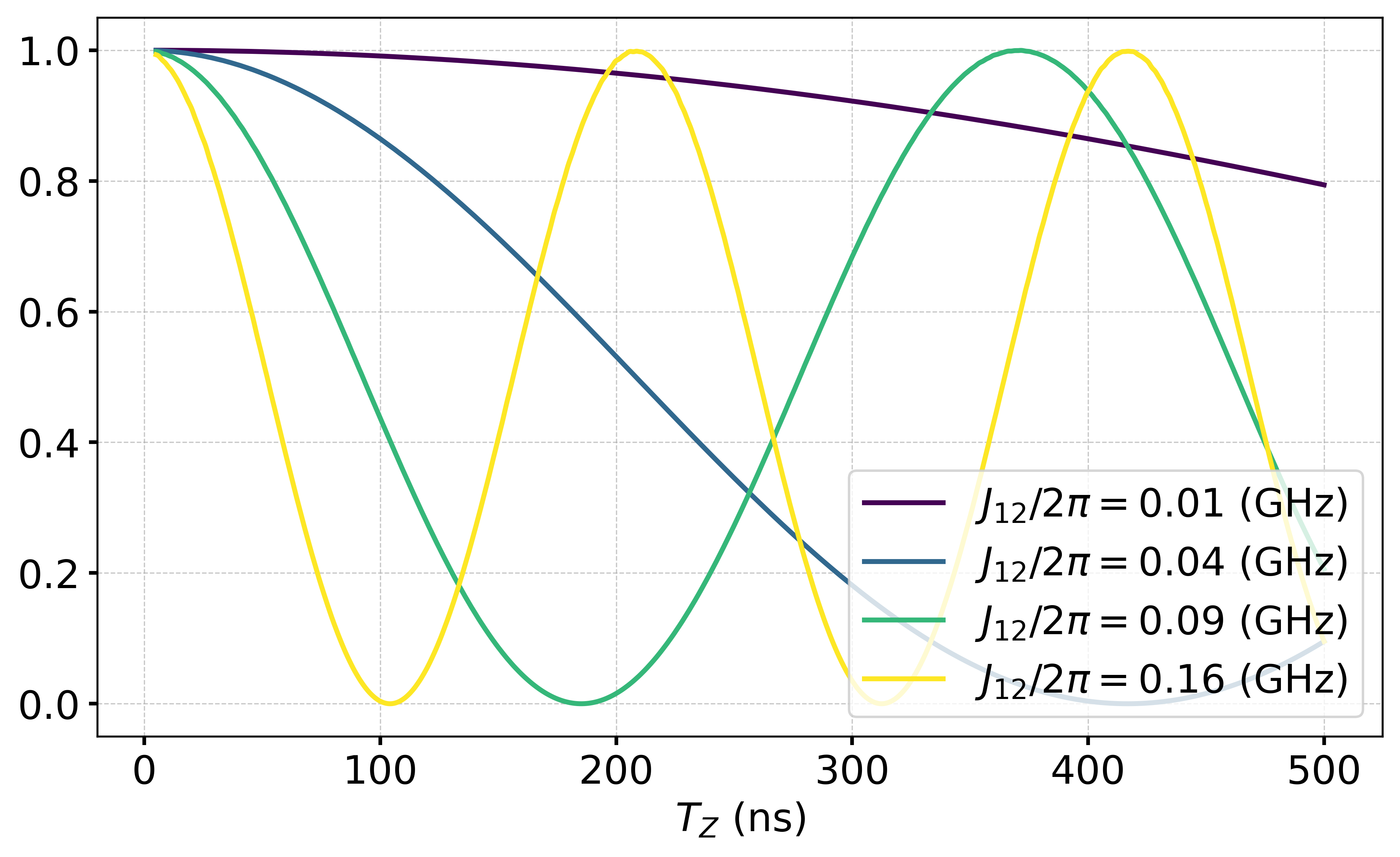}
        \label{fig: 3Q oscillation simple}
    \end{subfigure}
    \hfill
     % Subfigure (b)
    \begin{subfigure}{\linewidth}
        \centering
        \subcaption{\centering $|P_{DD}(\ket{+00},T_Z)-P(\ket{+00},T_Z)|$}
        \includegraphics[width=\linewidth]{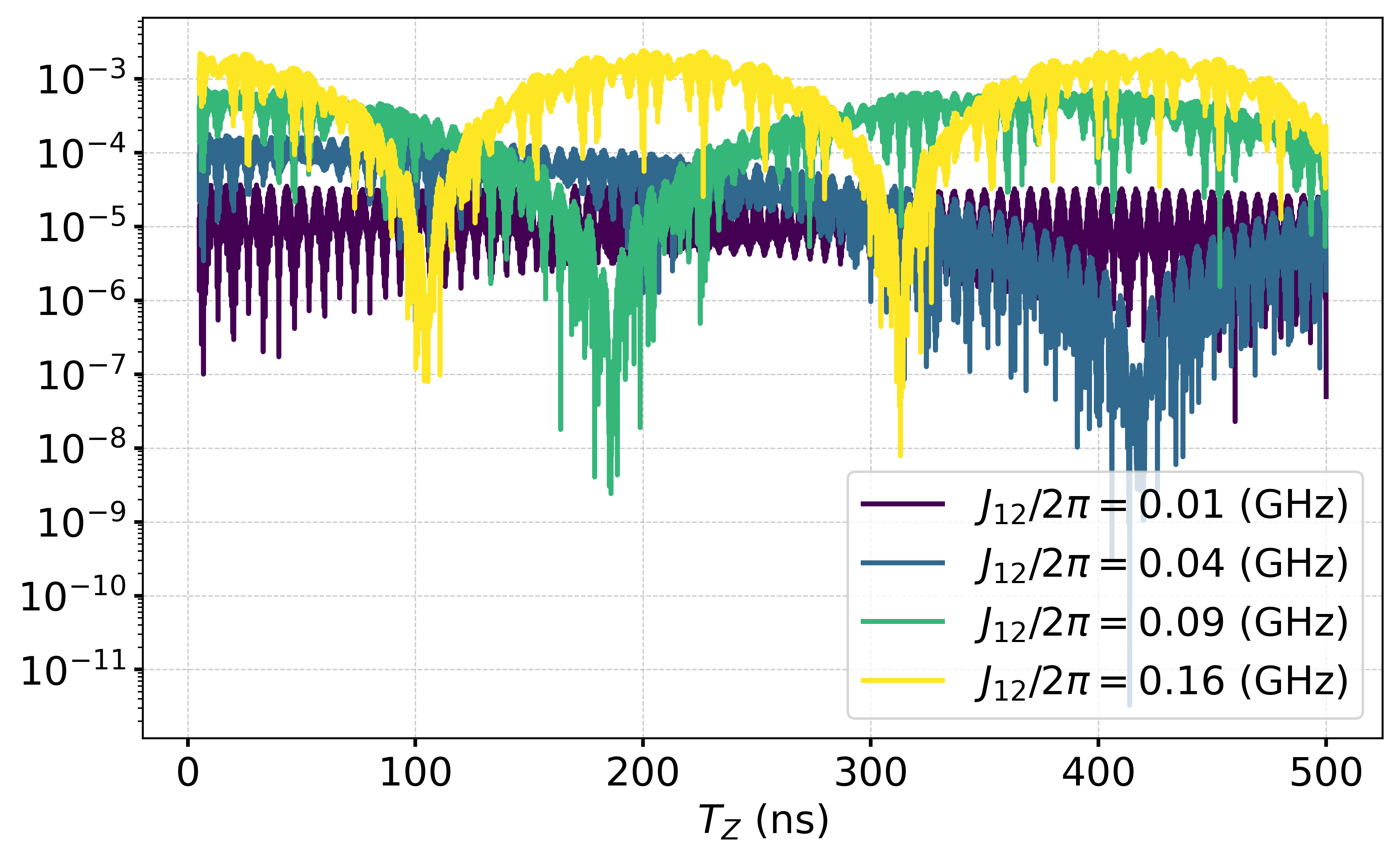}
        \label{fig: 3Q measurement error simple}
    \end{subfigure}%
    \caption{\justifying \textbf{Measurement of the chiral anisotropic exchange interaction using a 4-layer DD sequence.} We consider the full three-qubit Hamiltonian (including all flip-flop terms) for the optimal SOI configuration  of the $\rm C^2 Ry(\pi)$ gate (see Fig.~\ref{fig: SOI configuration plot}) with two-body exchange interaction strength $J_{12}/2\pi=J_{23}/2\pi$ given by $10~\rm MHz$ (Purple), $40~\rm MHz$ (Blue), $90~\rm MHz$ (Green), and $160~\rm MHz$ (Yellow).
    (a) Exchange oscillations induced by the three-qubit interaction $J_{123}^{\parallel}$, obtained from numerical simulations of the measurement circuit introduced in Eq.~\eqref{eq: J123 measurement circuit}. (b) Measurement error. Absolute difference between the expected probability curve, given by Eq.~\eqref{eq: expected probability curve}, and the numerically computed oscillation curve. Both plots are shown as a function of the free-evolution time $T_Z \in[5,500]~\rm ns$. Notably, at all considered $T_Z$, the deviation between the numerical and expected oscillation curves remains below $10^{-3}$.}
    \label{fig: 3Q measurement}
\end{figure}
We now discuss practical methods for measuring the $ZZZ$ interaction. One approach is to vary the external magnetic field $B_{ext}^{\perp}$ and perform spectroscopy measurements. If the three QDs form a closed loop, the energy spectrum will exhibit gap openings and closings as $B_{ext}^{\perp}$ is varied due to constructive and destructive interference effects, as a function of the amplitude $B_{ext}^{\perp}$. However, while spectroscopy can confirm the presence of a closed-loop system and chiral-symmetry breaking, it does not directly provide the interaction strength $J_{123}^{\parallel}$ nor can it demonstrate that the interaction is (strictly) of a $ZZZ$ nature. 

To quantitatively characterize $J_{123}^{\parallel}$, we design a dynamical decoupling (DD) protocol that selectively isolates the $ZZZ$ term from any generic three-qubit Hamiltonian  
\begin{equation}
    H_{\rm gen} = \sum_{i,j,k \in \{0,\dots,4 \} } \alpha_{ijk}~\sigma_{i} \sigma_{j} \sigma_{k},
\end{equation}
where $\{ \alpha_{ijk} \}$ are real coefficients. For simplicity, here we restrict ourselves to the analysis of a bang-bang sequence and leave a more detailed analysis of a bounded-control sequence \cite{Viola2003} for future work. 

For clarity, we provide a detailed description of the DD sequence and its optimal gate composition in Appendix.~\ref{appendix: dynamical decoupling sequence}. The DD sequence is applicable to all generic three-qubit Hamiltonians and consists of free evolution blocks interleaved with 32 layers of single-qubit gates, resulting in a total of 36 single-qubit gates. As a result, the effective Hamiltonian $H_{\rm DD }$ after the DD sequence is 
\begin{equation}
    H_{\rm DD } = \frac{J_{123}^\parallel}{8}~Z_1 Z_2 Z_3. 
\end{equation}
However, implementing the full DD sequence requires switching interactions on and off 32 times, which may be experimentally challenging. In Appendix~\ref{appendix: dynamical decoupling sequence}, we present an alternative, shorter sequence consisting of only four layers of single-qubit gates, totaling 8 single-qubit operations. Surprisingly, for the physically relevant Hamiltonian that implements the $\rm C^2 Ry$ gate, where the qubit frequency differences are large, numerical simulations show that this shorter DD sequence still provides accurate measurements. An explanation for this result is given in Appendix~\ref{appendix: dynamical decoupling sequence}.

Given the effective Hamiltonian $H_{\rm DD }$, the strength $J_{123}^\parallel$ can be measured by the following circuit 
\begin{equation}
\label{eq: J123 measurement circuit}
\begin{quantikz}
    \ket{+} & \ctrl{2}  &   \ctrl{1}  & \gate[3]{e^{-i H_{\rm DD} T_Z }} & \ctrl{2} & \ctrl{1} & \meter{X}  \\
    \ket{0} & &   \targ{X}  &       &  & \targ{X} &  \meter{Z}   \\
    \ket{0} & \targ{X} &     &       &  \targ{X} &  & \meter{Z}
\end{quantikz}
\end{equation}
By  reading out the oscillation frequency  of the probability  
\begin{equation}
    \label{eq: expected probability curve}
   P(\ket{+00},T_Z) = \cos^2\Big(\frac{J_{123}^{\parallel}}{8} T_Z  \Big) 
\end{equation}
of measuring $|+00\rangle$ as a function of $T_{Z}$, we can extract directly $J_{123}^\parallel$. 

We numerically verify that the 4-layer DD protocol can accurately detect the $ZZZ$ interaction with a high fidelity. The numerical performance of the more complex 32-layer DD protocol is provided in Appendix~\ref{appendix: dynamical decoupling sequence}. Here, we focus on the optimal SOI configuration for the $\rm C^2Ry(\pi)$ gate, presented in Fig.~\ref{fig: SOI configuration plot}, with $\theta_{13} =0,\gamma =\pi/2$, considering an ideal system with all flip-flop/flop-flop interactions included.

In Fig.~\ref{fig: 3Q oscillation simple}, we observe an oscillation curve that closely matches the theoretical prediction given by Eq.~\eqref{eq: expected probability curve}. Additionally, in Fig.~\ref{fig: 3Q measurement error simple}, we plot the absolute difference between the expected oscillation curve and the numerical simulation. We find that the discrepancy never exceeds a $10^{-3}$ error, even for strong two-qubit exchange coupling $J_{12}/2\pi = 160~\rm MHz$. This level of precision is sufficient to synchronize the direct three-qubit gates effectively. 

Moreover, we find that even with a miscalibration error of up to  15\% in $J_{123}^{\parallel}$, the $\rm C^{2}Ry(\pi)$ gate maintains a high fidelity, with $1-\bar{F} \in [10^{-4},10^{-3}]$. Therefore, we demonstrated that the DD protocol can accurately measure the $J_{123}^{\parallel}$ in an ideal system. In the next section, we investigate the performance of the $\rm C^{2}Ry(\pi)$ gate protocol under realistic noises.  

\section{Gate performance with  noise} 
\label{section: numerical results} 
\begin{figure*}
    \centering
    % First row: Quasi-static noise
    \begin{subfigure}{0.49\linewidth}
        \centering
        \caption{Fast gate under quasi-static error}
        \includegraphics[width=\linewidth]{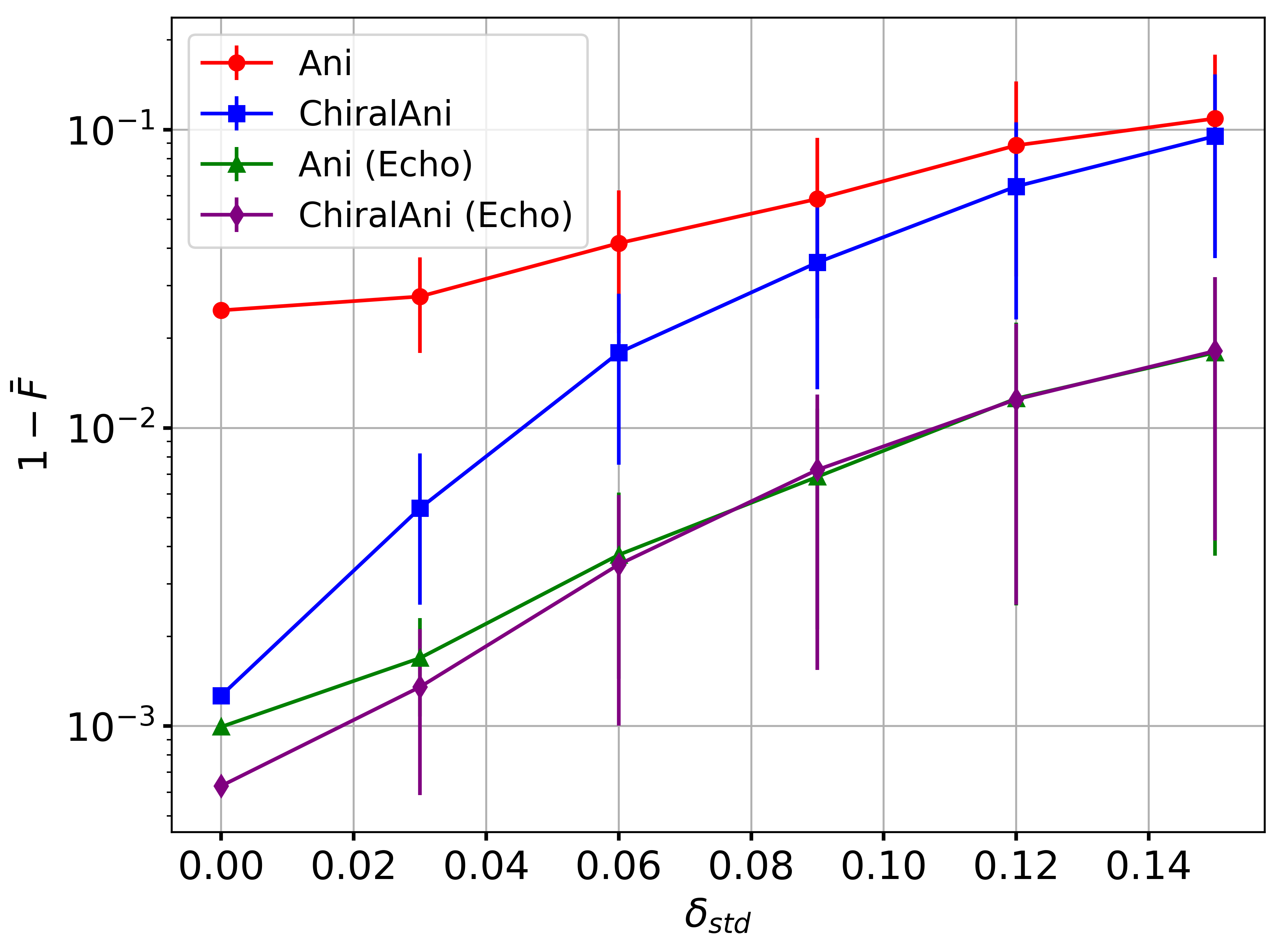}
        \label{fig: calib_fast}
    \end{subfigure}
    \hfill
    \begin{subfigure}{0.49\linewidth}
        \centering
        \caption{Normal gate under quasi-static error}
        \includegraphics[width=\linewidth]{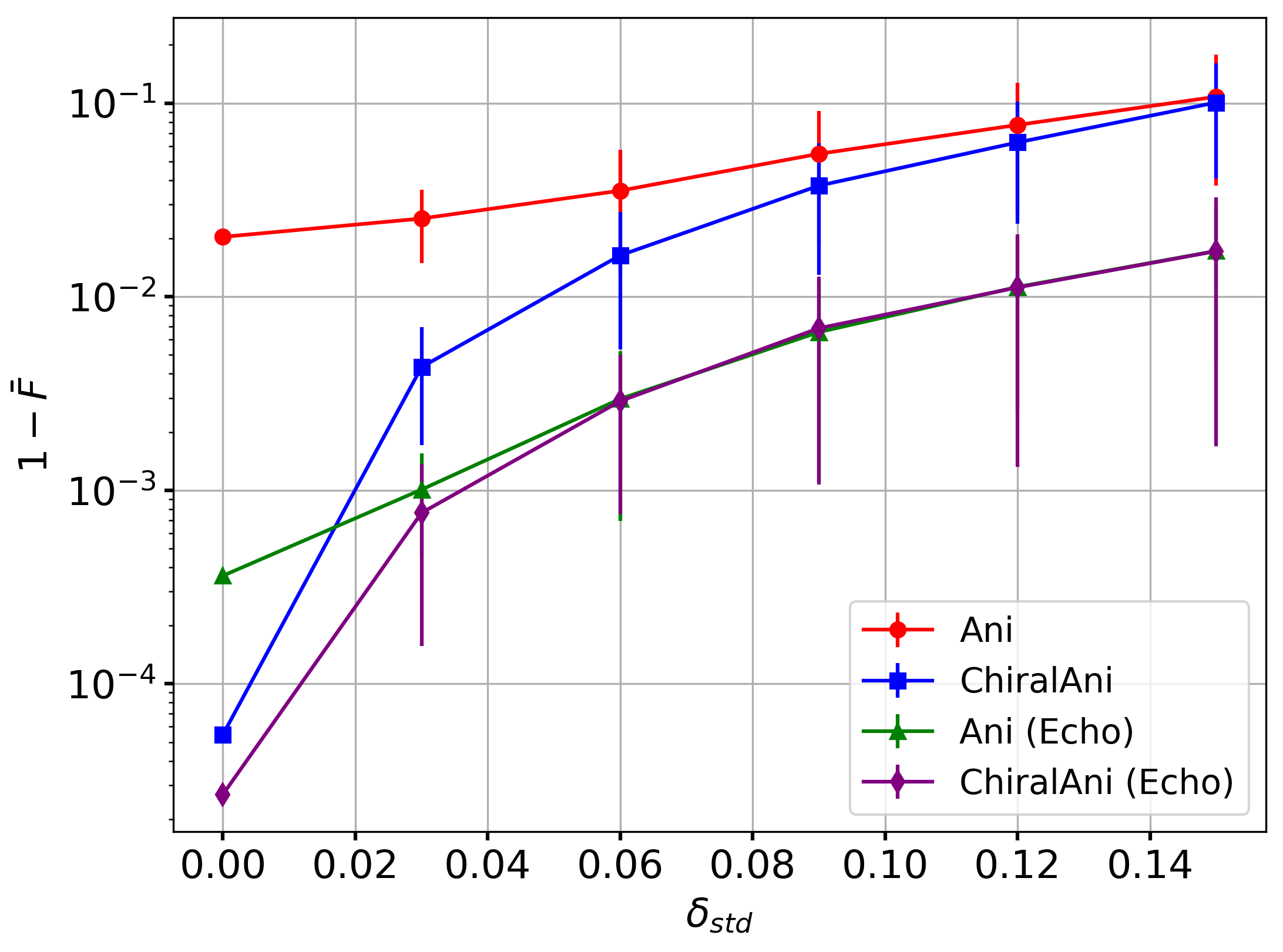}
        \label{fig: calib_slow}
    \end{subfigure}
    
    % Second row: 1/f noise
    \begin{subfigure}{0.49\linewidth}
        \centering
        \caption{Fast gate under $1/f$ noise}
        \includegraphics[width=\linewidth]{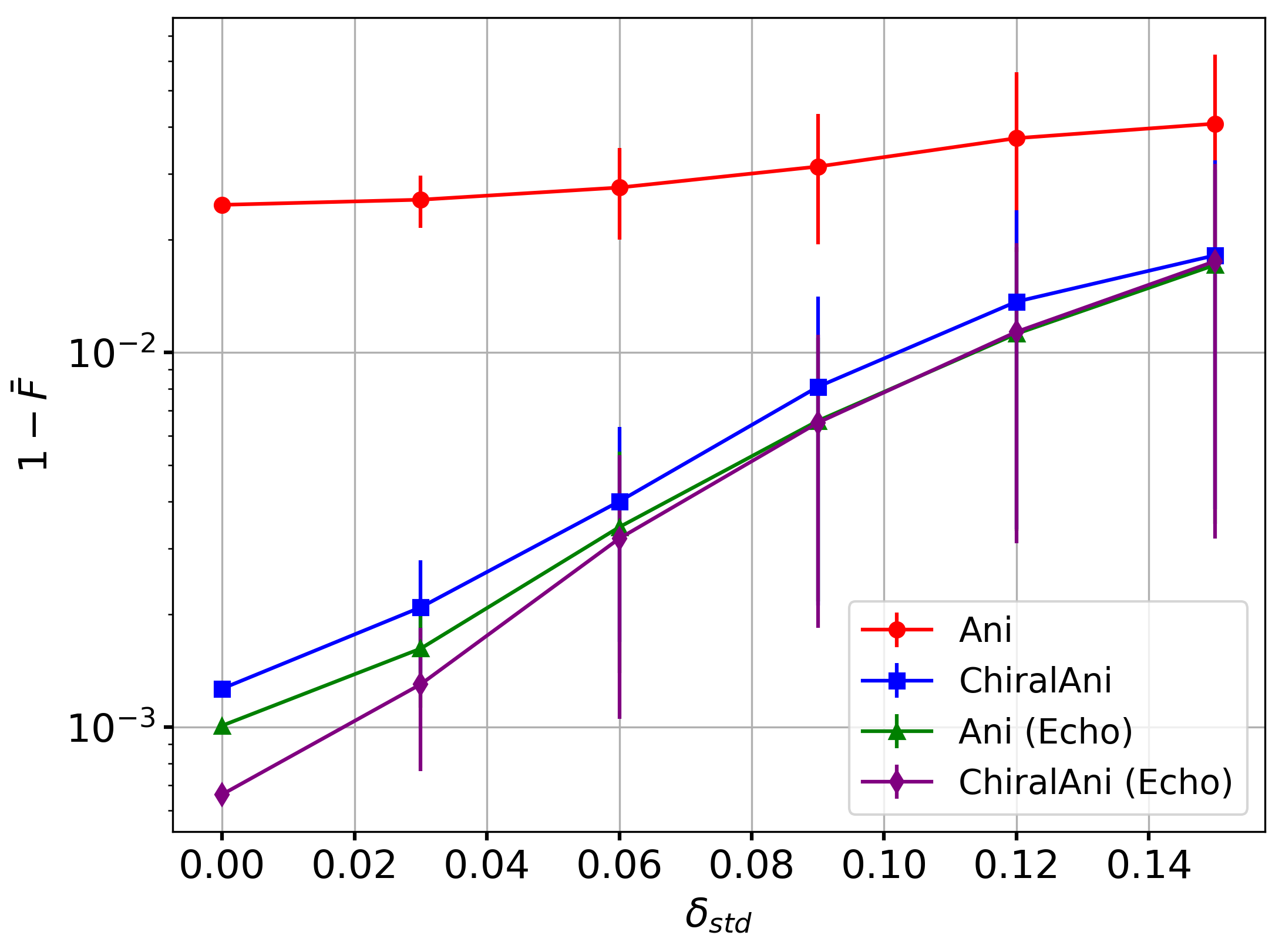}
        \label{fig: MonteCarlo_fast}
    \end{subfigure}
    \hfill
    \begin{subfigure}{0.49\linewidth}
        \centering
        \caption{Normal gate under $1/f$ noise}
        \includegraphics[width=\linewidth]{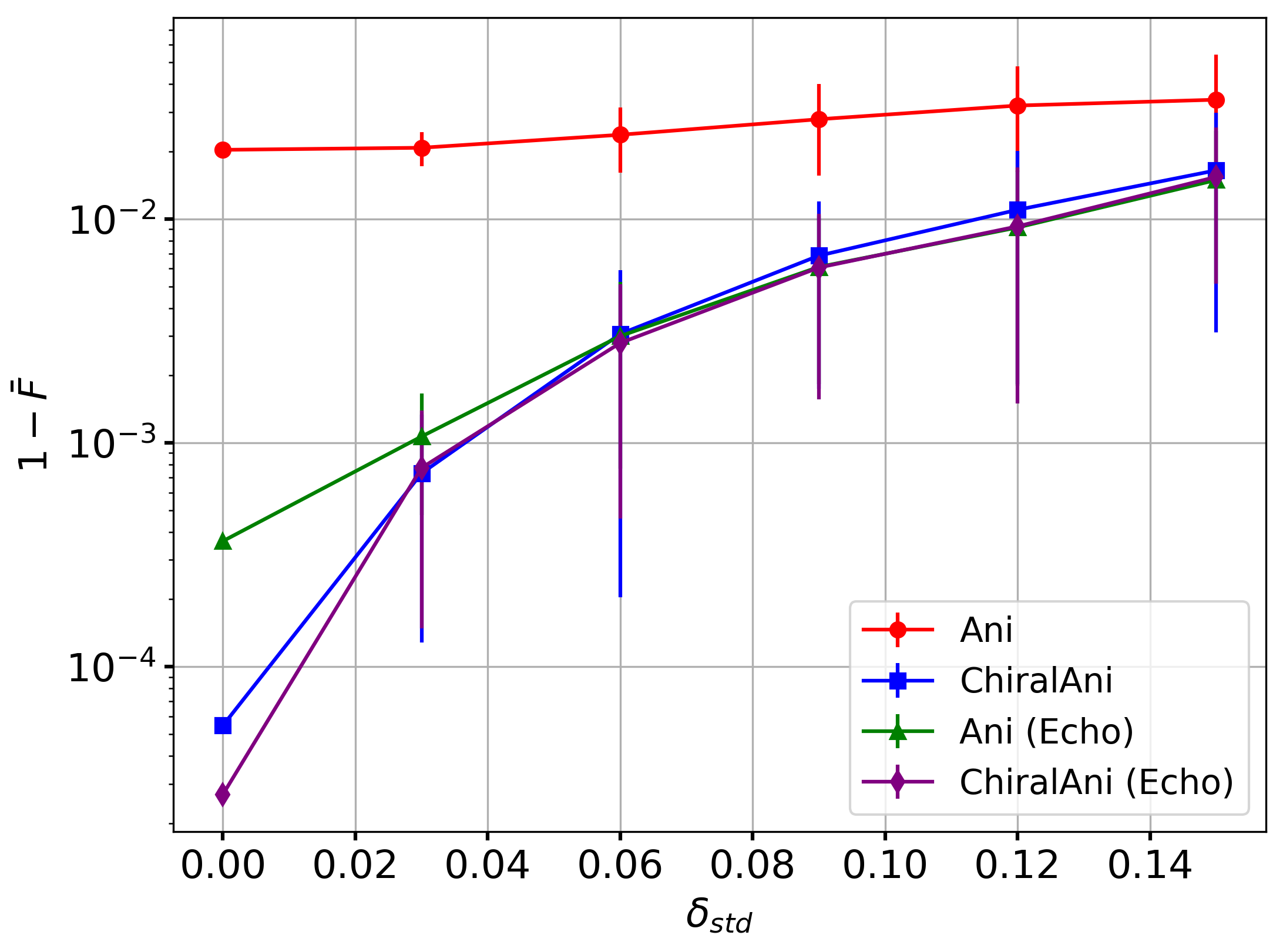}
        \label{fig: MonteCarlo_slow}
    \end{subfigure}
    
    \caption{\justifying \textbf{Gate performance with noise} Numerical simulation with qubits frequencies $\omega_1/2\pi = 4.2~\rm GHz$, $\omega_2/2\pi = 4.5~\rm GHz$, and $\omega_3/2\pi = 4.8~\rm GHz$. Exchange interactions are $J_{12}/2\pi = J_{23}/2\pi = 20~(90)~\rm MHz$ and $J_{13}/2\pi = 1.6~\rm MHz$ for normal (fast) gates with gate time $T_g = 100~(22.22)~\rm ns$. The Ani gate follows the $(0,1)$ synchronization, with Rabi frequencies $\Omega_2/2\pi \approx 5.16~(23.24)\rm MHz$, while the ChiralAni gate follows $(0,1,1,2)$, with $\Omega_2 /2\pi \approx 5.04~(22.68)\rm MHz$. The three-qubit interaction amplitude is $J_{123}^{\parallel}/2\pi \approx 0.96~(4.34)~\rm MHz$. We plot the gate infidelity under different noise sources for the anisotropic interaction (red), chiral anisotropic interaction (blue), and their Echo variants (green and purple, respectively). (a-b) Gate infidelity against quasi-static error, where the noise is modeled as $\delta_{i}(t) = \delta_i$, drawn from a uniform distribution $[-\delta_{std},\delta_{std}]$. (c-d) Gate infidelity under high-frequency $1/f$ noise, modeled as a white noise with fluctuation $\delta_{std}$ under infinite impulse response filter. Error bars represent $\pm$ one standard deviation from the mean.}
    \label{fig: MonteCarlo_combined}
\end{figure*}

In this section, we numerically evaluate the performance of the $\rm C^2Ry(\pi)$ gate implemented using the different schemes proposed in this work. We compare the Ani gate (which uses two-body interactions only) to the ChiralAni gate (which incorporates a three-body $ZZZ$ interaction), and with their Echo variants under simple noise models. For the definitions of these gates, we refer the reader back to Sec.~\ref{section: systematic errors}. More specifically, we include two noise sources: quasi-static errors and time-dependent $1/f$ noise, typical of hyperfine and charge noise. Since the gate-time $T_g$ is relatively fast, we assume that the $1/f$ noise also has finite weight at higher frequencies. 

The noisy Hamiltonian is given by
\begin{subequations}
    \begin{align}
         & H_{noisy}  = \sum_{i} \Big( \frac{\omega_i}{2} - \frac{ \delta_{123}(t) J_{123}^{\parallel}}{24}  \Big) Z_i \\
        & + \frac{1+\delta_{12}(t) }{4}J_{12}(X_1X_2-Y_1Y_2 + Z_1 Z_2) \\
        & + \frac{1+\delta_{23}(t) }{4}J_{23}(X_2 X_3-Y_2 Y_3 + Z_2 Z_3) \\
        & + \frac{1+\delta_{13}(t) }{4}J_{13}(-X_1 X_3 - Y_1 Y_3 + Z_1 Z_3) \\
        & + \frac{1+\delta_{123}(t) }{8}J_{123}^{\parallel}~  Z_1 Z_2 Z_3  ,
    \end{align}
\end{subequations}
where we include two-qubit flip-flop/flop-flop terms and qubit-frequency shifts induced by the chiral anisotropic interaction, see Eq.~\eqref{eq: exact chiral anisotropic interaction}. 

Physically, our noise model accounts for low-frequency fluctuations in tunneling amplitudes induced by variations in the barrier gate voltages $V_{ij}$. These fluctuations cause variations in both two-body exchange and three-body interaction terms, which can be comparable in magnitude relative to their respective strengths. Physically, these fluctuations can be attributed to charge noise, a critical noise source in semiconductor spin qubits~\cite{burkard2023semiconductor}. We adopt this model because tunnel coupling fluctuations are particularly detrimental: they shift the resonance frequencies Eq.~\eqref{eq: MW frequency} and affect the synchronization ratios Eqs.~\eqref{eq: synchronization ratios}. Although more realistic noise models could be considered, we believe that the one employed here effectively captures the essential features that influence gate performance under realistic conditions.

The noise terms $\delta_{i} (t) $ are treated as identically distributed random variables. For the quasistatic error, we assume $\delta_{i}(t)$ is time-independent, and drawn from a uniform distribution over the interval $[-\delta_{std},\delta_{std}]$. The choice of a uniform distribution is based on the assumption that noise is bounded by physical constraints, preventing rare but extreme fluctuations that can arise in a Gaussian model. This ensures that there is no unphysical outliers that could disproportionately affect the numerical average infidelity with small sample size. 

For the time-dependent noise model, $\delta(t)$ follows a time-correlated process with a power spectral density (PSD) given by $PSD(f) \propto 1/f$. We generate the $1/f$ noise with the same order of fluctuation $\delta_{std}$ as a white noise using the infinite impulse response filter (with 300 poles) described in Ref.~\cite{kasdin1995discrete}. We normalize the discrete $1/f$ noise such that on average it has the same power as white noises with fluctuation $\delta_{std}$ 
\begin{equation}
    \frac{\sum_{n=0}^{N_{\rm samples}} PSD(n f_s) }{N_{\rm samples}}= 2 \delta_{std}^2 \Delta t. 
\end{equation}
Because of the short gate time, our numerical method only includes high-frequency components (from $\sim 2 ~ \rm MHz$ up to $80~\rm GHz$) of the $1/f$ noise. The low-frequency components is partly captured by the quasi-static noise model.

For both noise models, we investigate numerically up to $\delta_{std} = 0.15$ with discrete step $\Delta \delta_{std}=0.03$. For each value of $\delta_{std}$, we sample $200$ realizations. This noise model is similar to the model considered in Refs.~\cite{Russ2018,Gullans2019,zajac2018resonantly,watson2018programmable}.

For the numerical simulations presented in Fig.~\ref{fig: MonteCarlo_combined}, we consider qubits with frequencies $\omega_1/2\pi =4.2~ \rm GHz,$ $\omega_2/2\pi = 4.5~ \rm GHz,$ and $\omega_3/2\pi=4.8~\rm GHz$. The exchange interactions for the normal (fast) gate are $J_{12}/2\pi = J_{23}/2\pi = 20 ~ \rm MHz~ (90~ \rm MHz)$ and $J_{13}/2\pi = 1.6~ \rm MHz$. For the Ani gate, we adopt the synchronization condition $(0,1)$, resulting in a Rabi frequency of $\Omega_2/2\pi \approx 5.16 ~\rm MHz$ for the normal gate and $\Omega_2/2\pi \approx 23.24 ~\rm MHz$ for the fast gate. Meanwhile, for the ChiralAni gate, we use the synchronization condition $(0,1,1,2)$, yielding $\Omega_2/2\pi \approx 5.04~\rm MHz$ for the normal gate and $\Omega_2 /2\pi\approx 22.68 ~\rm MHz$ for the fast gate. Correspondingly, the amplitude  of the three-qubit interaction is $J_{123}^{\parallel} /2\pi\approx 0.96~\rm MHz$ for the normal gate and $J_{123}^{\parallel} /2\pi\approx 4.34~\rm MHz$ for the fast gate. In Figs.~\ref{fig: calib_fast} and \ref{fig: calib_slow}, we observe that the ChiralAni gate fidelity decreases rapidly with increasing  $\delta_{std}$ in both operating regimes but has similar performance as the other gates when considering $1/f$ noise as shown in Figs.~\ref{fig: MonteCarlo_fast} and \ref{fig: MonteCarlo_slow}. However, we note that due to the discrete nature of our simulation, low-frequency components of $1/f$ noise manifest as quasi-static errors. Consequently, we find the ChiralAni gate fidelity rapidly converges to that of the Ani gate (Figs.~\ref{fig: calib_fast} and \ref{fig: calib_slow}). 

To investigate further, we conducted simulations where only the three-qubit interaction was subject to noise, revealing that the ChiralAni gate remains stable up to $\delta_{std}=0.15$. Therefore, we conclude that the fidelity drop is entirely due to stochastic fluctuations in two-qubit interactions. Since the ChiralAni gate already achieves high fidelity ($1-\bar{F} \leq 10^{-4}$) without noise, even small noise contributions become dominant and significantly impact performance. Nevertheless, even with a moderate amount of noise $\delta= 0.05$ (corresponding to $5\%$ fluctuations), the ChiralAni gate maintains an average fidelity of $\geq 99\%$. 

On the other hand, we observe in Fig.~\ref{fig: MonteCarlo_combined} that the four-step echo protocol (for both Ani and ChiralAni gates) is significantly more resilient to noise due to its echo structure. It typically outperforms the single-step protocol when moderate noise is present. A similar trend is seen in the ChiralAni (Echo) gate, where, due to its already high ideal fidelity, small two-qubit fluctuations become the dominant source of error. Consequently, the fidelity of the ChiralAni (Echo) gate converges to that of the Ani (Echo) gate. This trend is clearly visible for example in Fig.~\ref{fig: calib_fast}. These results suggest that the four-step protocol could be well-suited for a wide range of near-term devices, where fluctuations of two-qubit interactions remain moderate. The single-step ChiralAni protocol remains the optimal choice if the noise level is small. 

We also note, however, the four-step protocol also presents challenges not accounted for in our study, including the need for precise control over the exchange interaction during the echo pulse and accurately timing these pulses. Consequently, in terms of control overhead, our fully-synchronized  single-step protocol remains more appealing.

\section{Conclusion}
We introduced two new protocols for implementing a three-qubit controlled-controlled-rotation $\rm C^2Ry$ gate: (i) a single-step fully-synchronized protocol and (ii) a four-step protocol designed to further suppress systematic errors and noises. The protocols can be readily adapted to other three-qubit controlled-controlled-unitary gates. 

The single-step $\rm C^2 Ry$ gate protocol resolves the synchronization issue encountered in previous proposals by leveraging a small transverse three-qubit $ZZZ$ interaction. This interaction enables fast gate operation while maintaining perfect fidelity. Furthermore, we demonstrated that this interaction naturally arises in state-of-the-art spin qubit platforms through SOI and orbital magnetic fields. To facilitate experimental validation, we also proposed a measurement protocol based on dynamical decoupling that can accurately measure the $ZZZ$ interaction strength.  

In addition, we introduced a four-step protocol designed to implement a fast, high-fidelity $\rm C^2 Ry$ gate while mitigating systematic errors and noise. Through numerical simulations, we compared these protocols against recent state-of-the-art approach \cite{Gullans2019,Geyer2024,Rasmussen2020,Arias2021} for spin qubits and found that our methods could outperform existing protocols under realistic noise conditions and system parameters. Notably, the four-step protocol exhibits enhanced robustness against $1/f$ noise and quasi-static noise. These results contribute to scalable quantum computation in spin qubit platforms, overcoming circuit length limitation posed by low-fidelity multi-qubit gates.

\section{Acknowledgment}
This research was supported by the EU through the H2024 QLSI2 project,  by the Army Research Office under Award Number: W911NF-23-1-0110, and in part by NCCR Spin (grant number 225153). The views and conclusions contained in this document are those of the authors and should not be interpreted as representing the official policies, either expressed or implied, of the Army Research Office or the U.S. Government. The U.S. Government is authorized to reproduce and distribute reprints for Government purposes notwithstanding any copyright notation herein.

\bibliography{reference}

\appendix
\section{Infeasibility of controlled-controlled-rotation gate with two-qubit interactions}
\label{appendix: proof of infeasiblity}
In this appendix, we provide a proof of the infeasibility of the synchronization conditions for the more general $\rm C^2 Ry(\frac{p}{q}\pi)$ gate when only two-qubit interactions are available. The synchronization conditions for the $\rm C^{2} Ry \Big( \frac{p}{q} \pi \Big)$ gate are given by   
 \begin{subequations}
     \begin{align}
            \Omega_{2}~T_{g} & = \Big(4m+\frac{p}{q}\Big)\pi \\
            \sqrt{(J_{23}^{\parallel})^2 + \Omega_2^2 }~T_{g} & = 4n_1 \pi, \\
         \sqrt{(J_{12}^{\parallel})^2 + \Omega_2^2 }~T_{g} & = 4n_2 \pi, \\ 
         \sqrt{(J_{12}^{\parallel} + J_{23}^{\parallel})^2 + \Omega_2^2 }~T_{g} & = 4 n_3 \pi. 
     \end{align}
 \end{subequations}
Our main result is that if $\gcd(p,q) = 1$, the system of equations has no integer solutions $(m,n_1,n_2,n_3)$ unless $p/q$ is a multiple of four. In that case, $\rm C^{2} Ry \Big( \frac{p}{q} \pi \Big)$ gate reduces to the identity gate, and there are infinitely many integer solutions. \\
$\square$ \\ 
 We start the proof with the simplest case $q=1$, which we can restrict $p \in \{1,2,3\}$ since we can always absorb $\lfloor p/4 \rfloor$  into $m$. By rearranging the synchronization conditions, we find that integer solutions $\{m,n_1,n_2,n_3\}$ exists iff
\begin{subequations}
 \label{eq: reduced synchronization condition}
    \begin{align}
            & 2\sqrt{[(4n_1)^2 - (4m+p)^2][(4n_2)^2 - (4m+p)^2]} = \\
            &  (4n_3)^2+ (4m+p)^2 -(4n_1)^2 - (4n_2)^2 
    \end{align}
\end{subequations}
admits integer solutions. We prove this is impossible by contradiction. Suppose such integers exist for $p \in \{1,3\}$. Then, the RHS of Eq.~\eqref{eq: reduced synchronization condition} is an odd integer, while the left-hand side (LHS) is either irrational or even—leading to a contradiction. For $p=2$,  dividing both sides by two reduces the problem to the same contradiction.

For the general case where $q$ is any positive integer, suppose there exists an integer solution $\{m,n_1,n_2,n_3\}$ for the pair $(p,q)$. By repeating the gate $q$ times, there must also exist an integer tuple $\{m',n_1',n_2',n_3' \}$ for the pair $(p,1)$. However, we previously proved that no such solution exists when $(p \mod 4) \in \{1,2,3 \}$. Therefore, no integer solution exists for any positive integer pair $(p,q)$ with $\gcd(p,q)=1$ unless $p$ is a multiple of four. 

In the case where $p$ is a multiple of four, infinitely many non-trivial solutions exist when $q=1$, corresponding to the identity gate. For example, one such solution is $(m,n_1,n_2,n_3)= (5,7,7,11)$, satisfies the synchronization condition with $J^{\parallel}/\Omega = \sqrt{24/25}$. However, for $q>1$, no integer solutions exist under the realistic assumption that $n_1 = n_2 $. Since $\gcd(p,q)=1$, $q$  is not a multiple of four and we can rerwite $p=4 \tilde p $ with  $\gcd(\tilde p,q)=1$.  If an integer solution existed, it would require
\begin{equation} 
n_3^2 + 3\Big(m+\frac{\tilde{p}}{q}\Big)^2 - 4 n_1^2 = 0
\end{equation}
for some integers $\{m,n_1,n_3\}$. However, it is straightforward (but quite tedious) to show that the second term in this equation can never be an integer when $\gcd(\tilde p,q)=1$, which leads to a contradiction. This concludes the proof. \\
$\square$ 

We have established a no-go result for synchronizing general controlled-controlled-rotation gates. However, introducing a small longitudinal $ZZZ$ interaction resolves this issue, as explained in the main text. Given a synchronization condition $(m,n_1,n_2,n_3)$ for Eq.~\eqref{eq: synchronization condition 3Q}, determining the synchronization ratios $J_{ij}/\Omega_2$ is best done numerically due to the complexity of the analytical expressions. 

For the specific case $(m,n,n,2n)$, which is particularly useful as it ensures the three-body interaction remains an order of magnitude smaller than the two-body interaction, the synchronization ratios are explicitly given by:
\begin{subequations}
    \label{eq: (0nn2n) condition}
    \begin{equation}
        \frac{J^{\parallel}}{\Omega_2} = \frac{\sqrt{64 n^2-(4 m+1)^2}}{2 \sqrt{(4 m+1)^2}}, 
    \end{equation}
    \begin{equation}
        \frac{J_{123}^{\parallel}}{\Omega_2} = \frac{\sqrt{64 n^2-(4 m+1)^2}-2 \sqrt{16 n^2-(4 m+1)^2}}{\sqrt{(4 m+1)^2}}.
    \end{equation}
\end{subequations}
In the asymptotic limit of large $n$ and $m=0$, the ratio of the three-qubit to two-qubit interaction simplifies to
\begin{equation}
    \frac{J_{123}^{\parallel}}{J^{\parallel}} =  \frac{6 \tau_{13} \sin(\Phi_B) }{U} \approx \frac{3}{64 n^2}. 
\end{equation}
Therefore, the orbital magnetic field $B^{\perp}_{ext}$ can be arbitrarily small by increasing $n$, demonstrating that the protocol is highly tunable.  

\section{Microscopic model}
\label{appendix: miscroscopic model}
In this appendix, we provide additional details on the microscopic models used in our analysis and demonstrate the existence of the ESOA frame for a general single closed-loop system.

With the same assumptions of identical on-site potential $U_i =U$ and zero detuning $\epsilon_i=0$ as in the main text, an equivalent formulation of the Fermi-Hubbard Hamiltonian in Eq.~\eqref{eq: Fermi-Hubbard Hamiltonian} is given by
\begin{subequations}
    \label{eq: alternative FH model}
    \begin{align}
        H_{FH} & = \sum_{ \substack{\langle ij \rangle \\ \sigma \in \{\up,\dw\}} } (t_{ij,\sigma}~ \hat{c}_{j,\sigma}^{\dagger} \hat{c}_{i,\sigma} + s_{ij,\sigma} ~\hat{c}_{j,\bar{\sigma}}^{\dagger} \hat{c}_{i,\sigma}) \\
        & + \sum_{i} U_i~  \hat{n}_{i,\up } \hat{n}_{j,\dw}.
    \end{align}
\end{subequations} 
The tunneling is separated into two contributions $t_{ij,\sigma}$ and $s_{ij,\sigma}$. The notation $\bar{\sigma}$ denotes the opposite spin of $\sigma$. Importantly, while $t_{ij,\sigma}$ does not induce spin flips between $\ket{\up}$ and $\ket{\dw}$,  it still contributes to spin precession around the $Z$-axis. 

From the definition of the SOI unitary operator $\hat{S}_{rot}^{ij}$, the tunneling coefficients $\{ t_{ij,\sigma},s_{ij,\sigma} \}$ can be expressed in terms of the SOI vectors and angles as follows 
\begin{subequations}
    \label{eq: explicit tunneling coefficients}
    \begin{align}
        & t_{ij,\up} = \tau_{p}^{ij}-i (\tau_{s}^{ij} \cos \theta_{ij}),~t_{ij,\dw} = \tau_{p}^{ij} + i (\tau_{s}^{ij} \cos \theta_{ij}) \\
        & t_{ji,\up} = \tau_{p}^{ij} +i (\tau_{s}^{ij} \cos \theta_{ij}),~t_{ji,\dw} = \tau_{p}^{ij} - i (\tau_{s}^{ij} \cos \theta_{ij}) \\
        & s_{ij,\up} = -i\tau_{s}^{ij}   \sin(\theta_{ij})e^{-i \varphi_{ij}},~s_{ij,\dw} = - i \tau_{s}^{ij} \sin(\theta_{ij}) e^{i \varphi_{ij}} \\
        & s_{ji,\up} = i \tau_s^{ij} \sin(\theta_{ij})e^{-i \varphi_{ij}}, \quad s_{ji,\dw} = i \tau_s^{ij} \sin(\theta_{ij})e^{i \varphi_{ij}}. 
    \end{align}
\end{subequations}
Here, we separate the SOI contribution from the normal spin-preserving contribution
\begin{equation}
    \tau_{p}^{ij} = \tau_{ij} \cos(\gamma_{ij}),~ \tau_{s}^{ij} = \tau_{ij} \sin(\gamma_{ij}). 
\end{equation}
In the presence of a magnetic field $B_{\rm ext}^{\perp}$, we perform the Peierls substitution on the tunneling coefficients 
\begin{equation}
    t_{ij,\sigma} \to t_{ij,\sigma}~e^{i \phi_{B}^{ij}}, ~ s_{ij,\sigma} \to s_{ij,\sigma}~e^{i \phi_{B}^{ij}},
\end{equation}
where $\phi_{B}^{ij}$ is the Peierls phase obtained by the particle while traveling from site $i$ to site $j$.

Computing the $n$-th order effective Hamiltonian $H_{\rm eff,n}$ in Eq.\eqref{eq: t/U expansion} is a straightforward task. It involves evaluating all $n$-th order virtual tunneling events between the QDs and their associated tunneling amplitudes. These virtual tunneling processes are elegantly represented using the operators $T_{\pm1}$ and $T_{0}$, as introduced in Ref.~\cite{tUexpansion}. The matrix representations of these operators can be efficiently computed using Wick’s theorem.

Given the matrix representation of $H_{\rm eff,n}$, we can decompose it into Pauli operators. For instance, the Pauli-decomposition of the second-order effective Hamiltonian $H_{\rm eff,2}$ is given by 
\begin{widetext}
\begin{subequations}
\label{eq: full two-body exchange interaction}
\begin{equation}
    H_{\rm eff,2}  = \sum_{\langle ij \rangle}  \frac{J_{ij}}{4} \vec{\sigma}_j \cdot R(\gamma_{ij}, \theta_{ij},\varphi_{ij})\cdot \vec{\sigma}_i,
\end{equation}
    \begin{align}
    & R(\gamma_{ij},\theta_{ij},\varphi_{ij})  = \begin{bmatrix}
        \cos^2(\gamma_{ij}) & -\sin(2\gamma_{ij}) \cos(\theta_{ij}) & -\sin(2 \gamma_{ij}) \sin(\theta_{ij}) \sin(\varphi_{ij}) \\
        \sin(2\gamma_{ij}) \cos(\theta_{ij}) &  \cos^2(\gamma_{ij}) & -\sin(2 \gamma_{ij}) \sin(\theta_{ij}) \cos(\varphi_{ij}) \\
        \sin(2 \gamma_{ij}) \sin(\theta_{ij}) \sin(\varphi_{ij}) & \sin(2 \gamma_{ij}) \sin(\theta_{ij}) \cos(\varphi_{ij})  & \cos^2(\gamma_{ij})
    \end{bmatrix}\\
    & -\sin^2(\gamma_{ij})\begin{bmatrix}
        \cos^2(\theta_{ij}) - \cos(2 \varphi_{ij} ) \sin^2(\theta_{ij})  & \sin^2(\theta_{ij})\sin(2 \varphi_{ij}) & -\cos(\varphi_{ij})\sin(2 \theta_{ij}) \\
         \sin^2(\theta_{ij})\sin(2 \varphi_{ij}) & \cos^2(\theta_{ij}) + \cos(2 \varphi_{ij} ) \sin^2(\theta_{ij}) & \sin(2\theta_{ij}) \sin(\varphi_{ij}) \\
         -\cos(\varphi_{ij})\sin(2 \theta_{ij}) & \sin(2\theta_{ij}) \sin(\varphi_{ij}) & -\cos(2 \theta_{ij})
    \end{bmatrix}.
\end{align}
\end{subequations}
\end{widetext}

Without SOI $\gamma_{ij}=0$, we recover the standard Heisenberg exchange interaction $H_{\rm eff,2}^{iso}$, which is sufficient to perform universal quantum computation \cite{DiVincenzo2000}. However, the Heisenberg interaction contain off-diagonal flip-flop terms that are dangerous for resonant two-qubit gates and ultimately limit the speed of  high-fidelity controlled-rotations. This is because $H_{\rm eff,2}^{iso}$ couples the states $\{ \ket{i=0,j=1},\ket{i=1,j=0}\} $ which have a small energy gap $|\omega_i-\omega_j|$.  To perform a resonant controlled-rotation, we need to operate in the small exchange interaction regime $J_{ij} \ll |\omega_i-\omega_j| $ to avoid unwanted transitions due to the flip-flop terms $X_i X_j + Y_i Y_j$. As the Rabi frequency is proportional to the exchange strength through Eq.~\eqref{eq: synchronization ratios}, the gate speed is upper limited by the small qubit frequency difference $| \omega_i - \omega_j|$.

For a general third-order effective Hamiltonian $H_{\rm eff,3}$ without any symmetries, its Pauli decomposition is highly complex and involves a large number of Pauli terms. However, in the ESOA frame, two key symmetries significantly simplify the form of $\tilde{H}_{\rm eff,3}$: (i) all SOI angles are identical, $\gamma_{ij} = \gamma$, and (ii) all SOI vectors are identical and align along the $Z$-axis. These symmetries constrain the possible form of $\tilde{H}_{\rm eff,3}$, leading to a much simpler Pauli decomposition, as shown in Eq.~\eqref{eq: effective H2 in ESA frame}.

The diagonal elements of $\tilde{H}_{\rm eff,3}$ can be derived by considering all virtual tunneling events where the final state remains identical to the initial state. For a given initial state $\ket{\sigma \bar \sigma \bar \sigma}$, there are only two possible tunneling trajectories (along with their time-reversed counterparts), which are related by particle-hole symmetry:  
\begin{subequations}
    \begin{align}
        & \ket{\sigma \bar \sigma \bar \sigma} \rightarrow \ket{0, \sigma \bar \sigma, \bar \sigma} \rightarrow \ket{0,\bar \sigma,\sigma \bar \sigma} \rightarrow \ket{\sigma \bar \sigma \bar \sigma}, \\ 
        & \ket{\sigma \bar \sigma \bar \sigma} \rightarrow \ket{\sigma \bar \sigma, \bar \sigma, 0} \rightarrow \ket{\sigma \bar \sigma,0,\bar \sigma} \rightarrow \ket{\sigma \bar \sigma \bar \sigma}. 
    \end{align}
\end{subequations}
The accumulated spin-phase interference between these particle-hole trajectories gives rise to a prefactor of $\sin(3\tilde{\gamma})$, while interference between a trajectory and its time-reversed counterpart introduces an additional prefactor of $\sin(\Phi_B)$. This interplay between chiral symmetry breaking and SOI is the physical origin of the $ZZZ$ interaction in the chiral anisotropic interaction.

The explicit interaction strength $J_{123}^{\parallel}$ in the ESOA frame is found to be proportional to 
\begin{subequations}
    \label{eq: amplitude J123 from tunneling events}
    \begin{align}
        J_{123}^{\parallel} U^2 & \propto (t_{12,\up} t_{23,\up} t_{31,\up}-t_{12,\dw} t_{23,\dw} t_{31,\dw}) \\
        & + (t_{13,\up} t_{21,\up} t_{32,\up}-t_{13,\dw} t_{21,\dw} t_{32,\dw}),
    \end{align}
\end{subequations}
which we identify with third-order tunneling events. The first (second) term describes the spin-dependent phase interference effect between spin-up and spin-down particles traveling counterclockwise (clockwise) to the triangular device in Fig.~\ref{fig: triangular-device}. By using Eq.~\eqref{eq: explicit tunneling coefficients}, it is straightforward to find that the two terms are given by
\begin{subequations}
    \begin{align}
        (t_{12,\up} t_{23,\up} t_{31,\up}-t_{12,\dw} t_{23,\dw} t_{31,\dw}) & = -2i \tau^3  e^{i \Phi_B} \sin(3 \bar{\gamma}), \\
       (t_{13,\up} t_{21,\up} t_{32,\up}-t_{13,\dw} t_{21,\dw} t_{32,\dw}) & = 2i \tau^3 e^{-i \Phi_B}\sin(3 \bar{\gamma}).
    \end{align}
\end{subequations}
The two terms differ by a minus sign due to time-reversal symmetry. Consequently, the combination of magnetic phase interference and spin-dependent phase interference give rise to the prefactor in the main text 
\begin{equation}
    J_{123}^{\parallel} U^2 \propto 4 \tau^3  \sin(\Phi_B) \sin(3 \bar \gamma).
\end{equation}

To conclude this appendix, we provide a proof of the existence of the ESOA frame and extend our analysis to a general closed-loop system with $N \geq 3$ qubits.  In the main text, we assumed the existence of an average SOI operator 
\begin{equation}
    \hat{S}_{\rm ave} = \exp(-i \tilde{\gamma} \sigma_z)
\end{equation}
and a set of local transformation $\{ \hat{\mathcal{R}}_{i} \}$ such that 
\begin{equation}
    \hat{\mathcal{R}}^{\dagger}_{j} \hat{S}^{ij}_{\rm rot} \hat{\mathcal{R}}_i = \hat{S}_{\rm ave}. 
\end{equation}
We now prove that this condition can always be satisfied for any closed-loop system. 

$\square$ 

Due to the closed-loop connectivity, we consider the product of all SOI operators along the loop
\begin{equation}
    \hat{\mathcal{R}}_{1}^{\dagger}  \hat{S}^{N1}_{\rm rot} \dots    \hat{S}^{23}_{\rm rot}   \hat{S}^{12}_{\rm rot} \hat{\mathcal{R}}_{
    1},
\end{equation}
By performing cyclic permutations on this product, we obtain analogous expressions for other operators $\{ \hat{\mathcal{R}_{i}} \}$. Since cyclic permutations do not alter the energy spectrum, we define the ESOA frame by choosing $\{ \hat{\mathcal{R}}_{i} \}$  such that they diagonalize their corresponding product $\{\hat{S}^{ij}_{\rm rot}\}$.

We introduce the average SOI operator 
\begin{equation}
    \hat{S}_{\rm ave} = \exp(-i \tilde{\gamma} \sigma_z)
\end{equation}
and impose the condition 
\begin{equation}
    \label{eq: cyclic invariant SOI operator}
    \hat{S}_{\rm ave}^{N} = \hat{\mathcal{R}}_{1}^{\dagger}  \hat{S}^{N1}_{\rm rot} \dots  \hat{S}^{12}_{\rm rot} \hat{\mathcal{R}}_{
    1} = \hat{\mathcal{R}}_{2}^{\dagger}  \hat{S}^{12}_{\rm rot} \dots  \hat{S}^{23}_{\rm rot} \hat{\mathcal{R}}_{
    2} = \dots.
\end{equation}
By construction, the choice of $\{ \hat{\mathcal{R}}_{i} \}$ ensures that
\begin{equation}
    [\hat{\mathcal{R}}^{\dagger}_{j} \hat{S}^{ij}_{\rm rot} \hat{\mathcal{R}}_i,S_{ave}^{N}] = 0.
\end{equation}
This implies that in the ESOA frame, the transformed SOI operators are diagonal. To verify this, observe that cyclic invariant of Eq.~\eqref{eq: cyclic invariant SOI operator} enforces

\begin{equation}
    ( \hat{\mathcal{R}}_{
    j}^{ \dagger } \hat{S}^{ij}_{\rm rot} \hat{\mathcal{R}}_{
    i}) \hat{S}_{\rm ave}^{N} (\hat{\mathcal{R}}_{
    j}^{ \dagger } \hat{S}^{ij}_{\rm rot} \hat{\mathcal{R}}_{
    i})^{\dagger} = \hat{S}_{ave}^{N}.
\end{equation}
Thus, each transformed SOI operator commutes with $\hat{S}^{N}_{\rm ave}$, ensuring that they are diagonal in the ESOA frame.

Although we have established that the transformed SOI operators are diagonal, we now show that they are all equal to $\hat{S}_{\rm ave}$. This follows from a local $U(1)$ symmetry of the basis transformation. Specifically, the geometric mean $\hat{S}^{N}_{\rm ave}$ is invariant under the transformation
\begin{equation}
    \hat{\mathcal{R}}_{i} \to \hat{\mathcal{R}}_{i} e^{-i \alpha_i \sigma_{z}}. 
\end{equation}
By appropriately choosing $\{ \alpha_i \}$, we can change the rotation angles of all transformed SOI operators such that
\begin{equation}
    \hat{\mathcal{R}}^{\dagger}_{j} \hat{S}^{ij}_{\rm rot} \hat{\mathcal{R}}_i = \hat{S}_{\rm ave} = \exp(-i \tilde \gamma \sigma_z)
\end{equation}
This completes the proof. 
$\square$

\section{Generalization of echo protocol to multi-controlled gates}
\label{appendix: generalization to N-qubit}
In the main text, our analysis has primarily focused on the three-qubit controlled-controlled-rotation gate. However, our results naturally extend to general multi-control single-target $\rm C^{N-1}Ry$ gates. A key insight is that achieving fast, high-fidelity $\rm C^{N-1}Ry$ gate in a single step requires all-to-all connectivity between the control qubits and the target qubit to perfectly synchronize higher-order off-resonant transitions \cite{Jiaan2024}. 

In spin qubits, this condition translates to the need for increasingly higher-order interactions, which become challenging to control. Specifically, there are  $2^{N-1}$ resonant transitions that must be synchronized, requiring an equal number of tunable parameters. While one degree of freedom is provided by the Rabi frequency of the driving microwave, the remaining $2^{N-1}-1$ parameters must come from all possible $Z$-type $k$-qubit interactions between the set of control qubits and the target qubit.
 \begin{equation}
     \sum_{k=1}^{N-1} \binom{N-1}{k} = 2^{N-1}-1.
 \end{equation}
For example, implementing a perfect $\rm C^3 Ry$ gate in single-step requires one Rabi frequency, three two-qubit $Z_{i}Z_{j}$ interactions, three three-qubit $Z_{i} Z_{j} Z_{k}$ interactions, and one four-qubit $Z_{1} Z_{2} Z_{3} Z_4$ interaction.

\begin{figure}
    \centering
    \includegraphics[width=\linewidth]{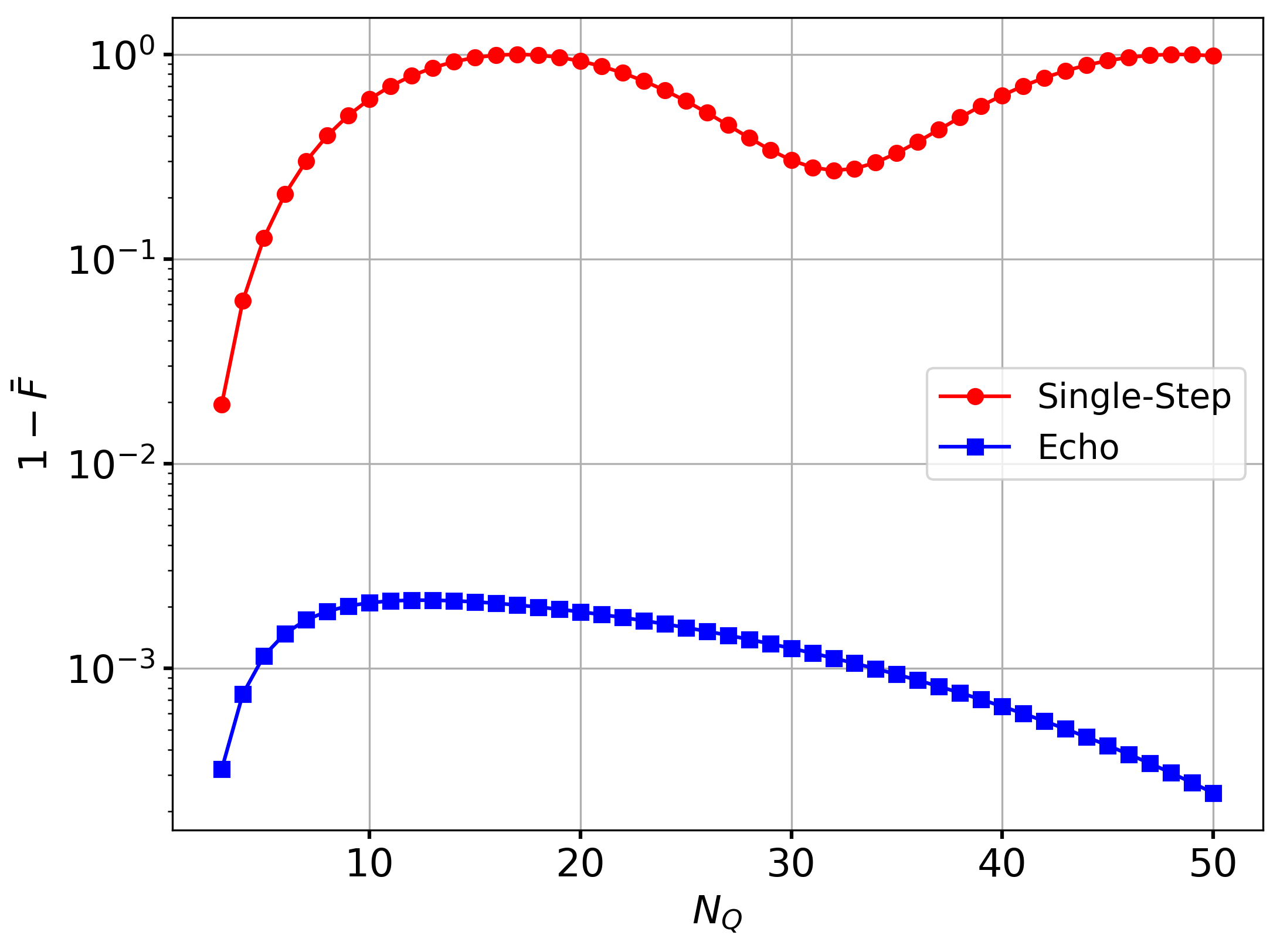}
    \caption{\justifying \textbf{Multi-controlled gate fidelity.} Average fidelity of the $\rm C^{N_Q} Ry$ gate as a function of the number of control qubits $N_Q$, implemented using the single-step protocol (red) and the four-step protocol (blue). The number of control qubit starts as $N_Q = 2$ and end at $N_Q=50$. }
    \label{fig: Multi-Q-infidelity}
\end{figure}

Furthermore, this analysis suggests that the fidelity of the multi-control $\rm C^{N-1} Ry$ gate, when implemented using only two-qubit interactions, will deteriorate rapidly as the number of control qubits increases. This degradation arises from the exponential scaling of off-resonant transitions with the number of control qubits. Assuming uniform qubit-qubit interactions $J^{\parallel}$ and applying single-qubit phase corrections, we can analytically derive the average gate fidelity of the $ \rm C^{N-1} Ry$ gate as a function of the number of qubits $N$:
\begin{subequations}
    \begin{align}
    \label{eq: multi-control average fidelity}
    & \bar{F}= \frac{1}{2^{N}+1} +\\
    & \frac{ 2^{N} }{2^{N}+1} \Big( \frac{1}{2^{N-1}}+ \sum_{k=1}^{N-1} \frac{\binom{N-1}{k}}{2^{N-1}} \cos[ \frac{\sqrt{(kJ^{\parallel}/\Omega)^2+1} \pi}{2}  ]  \Big)^2  .
    \end{align}
\end{subequations}
In Fig.~\ref{fig: Multi-Q-infidelity}, we plot the average fidelity of the $\rm C^{N_Q} Ry$ gate as a function of the number of control qubits $N_Q=N-1$ (starting at $N_Q=2$), under the synchronization condition $J^{\parallel} = \sqrt{15} \Omega$. The results show that the gate fidelity drops below $90\%$ already at $N_Q = 4$, corresponding to the five-qubit $\rm C^{4}Ry$ gate. On the other hand, if we employ the four-step protocol of the gate, the average gate fidelity is given by

\begin{subequations}
    \begin{align}
     & \bar{F}_{\rm echo}= \frac{1}{2^{N+1}} +\frac{2^{N}}{2^{N}+1}\times \\
           & \Big[ \frac{1}{2^{N-1}} + \sum_{k=1}^{N-1} \frac{\binom{N-1}{k} }{2^{N-1}}\frac{(kJ^{\parallel})^2 + \Omega^2  \cos(\frac{\pi}{2} \sqrt{(\frac{k J^{\parallel}}{\Omega} )^2 +1})}{(k J^{\parallel})^2 + \Omega^2}  \Big]^2 .
    \end{align}
\end{subequations}

In Fig.~\ref{fig: Multi-Q-infidelity}, we plot the average gate fidelity of the four-step protocol  under the same synchronization condition $J^{\parallel} = \sqrt{15} \Omega$ as the single-step protocol. We observe that the four-step protocol mitigates errors much more effectively than the single-step protocol due to its echo structures, which helps suppressing high-order off-resonant transitions. Notably, the four-step protocol fidelity increases as the number of control qubits increases. This counterintuitive effect arises because the dominant source of error in the four-step protocol is second-order off-resonant transitions ($k=2$) whose number scales quadratically with $N_Q$, whereas the total number of transitions scales exponentially \cite{Rasmussen2020}. This scaling imbalance leads to an overall improvement in fidelity as $N_Q$ increases.

Even in the worst-case scenario, where the state resides in a second-order off-resonant transition subspace, our results show that the four-step protocol can still achieve four-nines fidelity while requiring only twice the gate time of a standard two-qubit gate. Furthermore, as demonstrated in the main text, introducing a three-qubit coupling term $Z_i Z_j Z_k$ can further suppress errors, extending error mitigation to third-order off-resonant transitions. This suggests that by combining the four-step protocol with the chiral anisotropic interaction, it will be possible to implement a fast and high-fidelity four-qubit gate in near-future devices.

\section{Dynamical decoupling sequence for measurement}
\label{appendix: dynamical decoupling sequence}
In this appendix, we briefly review the basic of dynamical decoupling using group theory. The task of designing a DD sequence can be viewed as choosing a subgroup $G$ of the Pauli group and twirling the Hamiltonian $H_{\rm gen}$ over the element of group $G$, namely 
\begin{equation}
    \frac{1}{|G|} \sum_{\hat{g} \in G} \hat{g}H_{\rm gen}\hat{g}^{\dagger}. 
\end{equation}
The twirling operation projects the $H_{\rm gen}$ onto the subgroup of Pauli matrix which commutes with the group $G$. 

We will use a two-level concatenation DD scheme \cite{Lidar2002} to isolate the $ZZZ$ interaction in any generic three-qubit Hamiltonian 
\begin{equation}
    H_{\rm gen} = \sum_{i,j,k \in \{0,\dots,4 \} } \alpha_{ijk}~\sigma_{i} ~\sigma_{j} ~\sigma_{k}.
\end{equation}
In the first level, we project the generic Hamiltonian $H_{\rm gen}$ onto its diagonal elements by twirling over the $Z$-group $G_1$ generated by
\begin{equation}
    G_1 = \langle Z_{1},Z_{2}, Z_{3} \rangle,
\end{equation}
since the only Pauli matrices commute with $G_1$ are $Z$-type Pauli strings. In the second level, we twirl over the $X$-group $G_2$ generated by
\begin{equation}
    G_2 = \langle X_{1}X_{2},X_{2} X_{3} \rangle. 
\end{equation}
We can easily verify that all one-qubit and two-qubit Z-type Pauli strings anti-commute with at least one of $G_2$'s generators. Therefore, after the DD sequence, we are left with an effective Hamiltonian containing only the $ZZZ$ interaction. 
\begin{figure}
    \centering
     % Subfigure (b)
    \includegraphics[width=\linewidth]{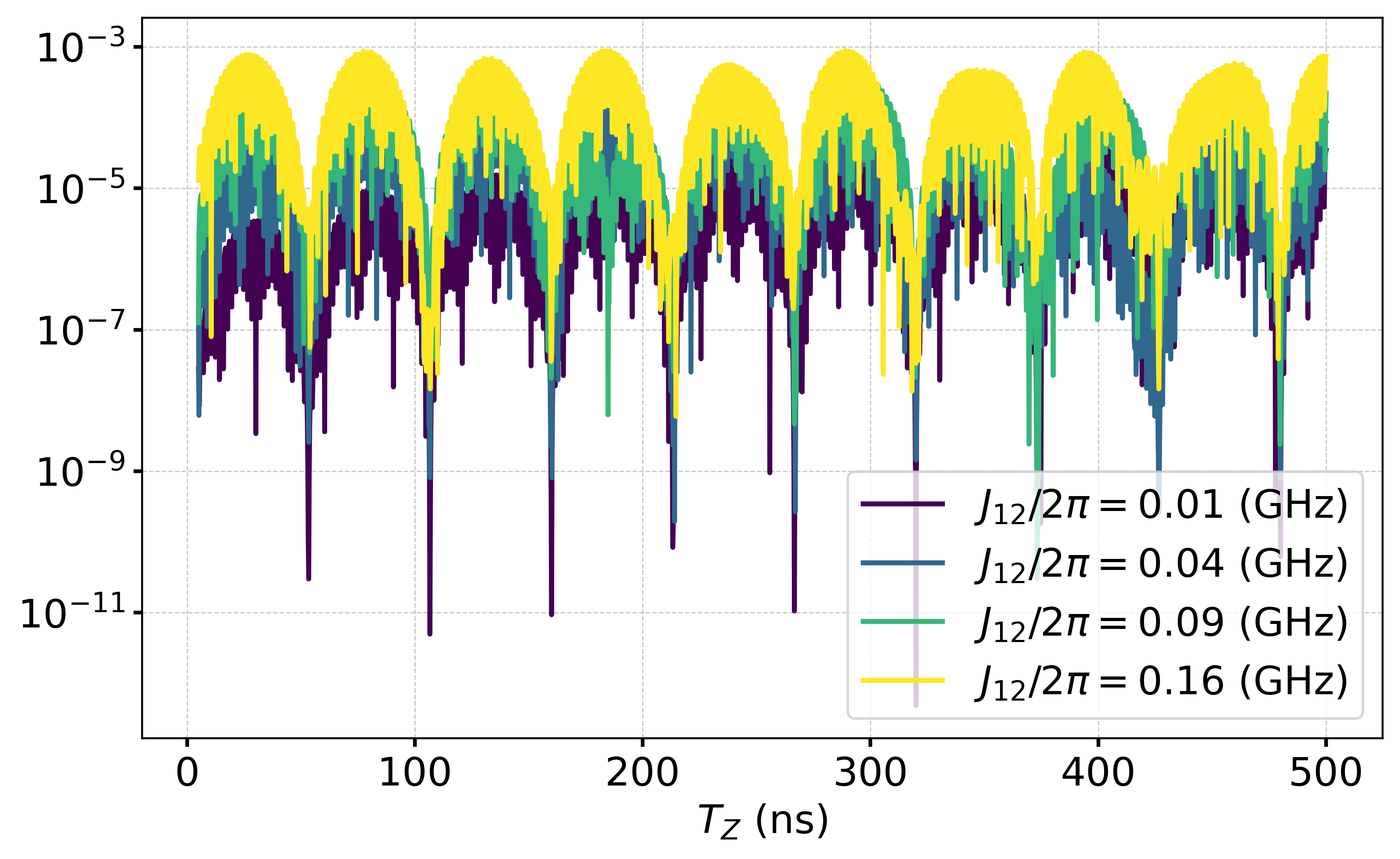}
    \caption{\justifying  \textbf{Measurement error of the chiral anisotropic exchange interaction using the 32-layer DD sequence.}  We consider full three-qubit Hamiltonian (including all flip-flop terms) for the optimal SOI configuration of the $\rm C^2 Ry(\pi)$ gate with two-body exchange interaction strength $J_{12}/2\pi=J_{23}/2\pi$ given by $10~\rm MHz$ (Purple), $40~\rm MHz$ (Blue), $90~\rm MHz$ (Green), and $160~\rm MHz$ (Yellow).}
    \label{fig: 3Q measurement complex}
\end{figure}

By construction, the above sequence consists of $|G_1| \times |G_2| =32 $ layers of single-qubit gates. However, we can also estimate the minimum number of single-qubit gates required to implement the DD sequence. First, the optimal decomposition for twirling over the group $G_1$ can be determined by following a Hamiltonian path starting at identity in the Cayley graph of $G_1$. This requires $8$ single-qubit gates. When concatenating $G_1$ and $G_2$, each iteration over $G_1$ needs one additional single-qubit gates, come from the elements of $G_2$. As a result, the minimum number of single-qubit gates required for the DD sequence is $36$. 

Instead of applying the full concatenated DD scheme over $G_1 \times G_2$, we define a simpler four-layer DD sequence by twirling only over the group $G_2$.  The $4$-layer DD sequence requires only $8$ single qubits gate under optimal decomposition. As we will discuss shortly, this approach significantly reduces both the number of applied gates and the number of layers, minimizing the number of times the interaction needs to be turned on and off. Despite this simplification, the four-layer DD sequence retains high accuracy when applied to physically relevant Hamiltonians, particularly those with large Zeeman energy splitting and significant qubit frequency differences.

In Fig.~\ref{fig: 3Q measurement complex}, we show the absolute difference between the expected oscillation curve in Eq.~\eqref{eq: expected probability curve} and the numerical simulation using the 32-layer DD sequence for the optimal SOI configuration of the $\rm C^2Ry(\pi)$ gate. Notably, the performance of the 4-layer DD sequence, shown in Fig.~\ref{fig: 3Q measurement error simple}, is comparable to that of the 32-layer DD sequence, with absolute errors remaining below $10^{-3}$ for all run time $T_Z$. 

The  high accuracy of the 4-layer DD sequence can be attributed to the large Zeeman energy splitting and significant qubit frequency differences, both of which naturally suppress flip-flop interactions, effectively mimicking the effect of twirling over the group $G_1$. Alternatively, we can understand the high accuracy considering that to first-order SW transformation, the transformed Hamiltonian is Ising, with off-diagonal elements appearing as small phase-shift. Consequently, we only need to twirl over the group $G_2$. 

Nevertheless, we can see a clear difference between the 32-layer DD sequence and the 4-layer DD sequence by comparing the shape of the error curve. Specifically, the error curve of the 4-layer DD sequence exhibits a dependence on the three-qubit interaction strength $J_{123}$, whereas the error curve of the 32-layer DD sequence does not. This difference arises from the greater leakage out of the $\{ \ket{000},\ket{100} \}$  subspace in the 4-layer DD sequence compared to the 32-layer DD sequence.

\end{document}